\documentclass[12pt,a4paper]{article}
\usepackage{amsfonts}
\usepackage{amssymb}
\usepackage{graphicx}
\usepackage{amsmath}
\usepackage{makeidx}
\usepackage{indentfirst}
\usepackage[T1]{fontenc}
\usepackage[utf8]{inputenc}
\usepackage{authblk}

\newcounter{resultnum}[section]
\newcounter{conclusionnum}[section]
\newcounter{conditionnum}[section]
\newcounter{conjecturenum}[section]
\newcounter{examplenum}[section]
\newcounter{exercisenum}[section]
\newcounter{lemmanum}[section]
\newcounter{notationnum}[section]
\newcounter{theoremnum}[section]
\newcounter{definitionnum}[section]
\newcounter{corollarynum}[section]
\newcounter{remarknum}[section]
\newcounter{propositionnum}[section]
\newcounter{acknowledgementnum}[section]
\newcounter{algorithmnum}[section]
\newcounter{axiomnum}[section]
\newcounter{casenum}[section]
\newcounter{claimnum}[section]
\newcounter{summarynum}[section]
\newcounter{problemnum}[section]
\begin{document}

\title{Modified Dynamical Supergravity Breaking and Off--Diagonal
Super-Higgs Effects}
\date{October 2, 2014}

\author[1]{Tamara  Gheorghiu \thanks{tamara.gheorghiu@yahoo.com}}%

\author[2]{Olivia Vacaru\thanks{olivia.vacaru@yahoo.com}}%

\author[3]{Sergiu  Vacaru\thanks{sergiu.vacaru@cern.ch;\ sergiu.vacaru@uaic.ro}}%

\affil[1]{\small Project IDEI, Alexandru Ioan Cuza University,\ Alexandru Lapu\c sneanu street, \newline  nr. 14, UAIC - Corpus R, office 323;
\  Ia\c si,\ Romania, 700057
\newline {\qquad }
}

\affil[2]{\small National College of Ia\c si;\ Arcu street, nr. 4, Ia\c si,\ Romania,\ 700125
  \newline {\qquad }
}

\affil[3]{\small Theory Division, CERN, CH-1211, Geneva 23, Switzerland \footnote{associated visiting researcher};\ and
\newline
 Rector's Office, Alexandru Ioan Cuza University,\
 Alexandru Lapu\c sneanu street, \newline  nr. 14, UAIC - Corpus R, office 323;\
 Ia\c si,\ Romania, 700057
}%

\maketitle

\begin{abstract}
We argue that generic off--diagonal vacuum and nonvacuum solutions for
Einstein manifolds mimic physical effects in modified gravity theories
(MGTs) and encode certain models of the $f(R,T,...)$, Ho\v rava type with
dynamical Lorentz symmetry breaking, induced effective mass for the graviton
etc. Our main goal is to investigate the dynamical breaking of local
supersymmetry determined by off--diagonal solutions in MGTs and encoded as
effective Einstein spaces. This includes the Deser-Zumino super--Higgs
effect, for instance, for a one--loop potential in a (simple but
representative) model of $\mathcal{N}=1, D=4$ supergravity. We develop and
apply new geometrical techniques which allows us to decouple the
gravitational field equations and integrate them in a very general form with the
metric and vielbein fields depending on all the spacetime coordinates via means of
various generating and integration functions and parameters. We study how
solutions in MGTs may be related to the dynamical generation of a gravitino mass
and supersymmetry breaking.

\vskip0.1cm


\vskip3pt

\end{abstract}

\tableofcontents

\renewcommand\Authands{
and  }



\section{Introduction}

Recently, the Nobel Prize in Physics 2013 was awarded to F. Englert and P.
W. Higgs \cite{nbp}. Their theoretical work was "confirmed'' after the
discovery of the predicted fundamental particle, by the \ ATLAS and CMS
experiments at CERN's Large Hadron Collider" \cite{atlas}. Related to
this discovery, we note that supersymmetry remains still an important theoretical activity
in the construction of various unifying models
in low energy physics.  For instance, in the study of the stability of the Higgs vacuum and
unification of the fundamental forces, and/or the low scale of inflation.
Supergravity (SG) theories are conjectured to be related to the low--energy
limits of superstring theories and which are the candidates for a theory of quantum gravity
and its unfication with the other fundamental interactions.

If it exists, supersymmetry must be broken at low energies because
there is not any experimental data (and related phenomenological models)
confirming such generalized symmetry. We cite early work \cite{supersymy}
when the general relativity (GR) theory was obtained in the standard low energy limit
of superstring and supegravity theory. Perhaps this is related to the
spontaneous or dynamical symetry breaking of a fundamental theory
and to various modified gravity theories (MGTs). We consider that it is desirable to elaborate a formalism of breaking the local supersymmetry directly, i.e. without the coupling to a gauge sector. A
typical example is the one used for the breaking of chiral symmetry in \cite{nambu}%
. Our goal is to extend this approach when MGTs (and in particular,
off--diagonal solutions in GR) are included into the scheme of dynamical
breaking of local supersymmetry.

There are several important motivations for
the elaboration and study of  MGTs. A number of theoretical attempts
have been proposed in order to explain the accelerated expansion of the
universe, solve the dark energy and dark matter problems and find consistency with
the observational data in modern cosmology.   To formulate a
self--consistent quantum gravity theory seems very difficult  to achieve within the
framework of the ordinary GR theory.  We cite here some of most popular approaches
elaborated during the last 25 years. For instance, the DGP braneworld model by
Dvali, Gabadadze and Porrati \cite{dgp}) is a version of the theories with
warped/ trapped configurations and extra dimensions  which can be
extended to off--diagonal configurations, see \cite{voffds}. There is
certain "cosmological" equivalence between such models and some classes of
Finsler-like theories \cite{bs}, see papers \cite{vrevlf} for reviews and
recent results on the so--called Lagrange--Finsler modifications of gravity
theories. Perhaps, one of the most popular class of MGTs consists of the
so--called $f$--modified theories when certain functionals $\ f(R),f(R,T)$ of the curvature and torsion
modify the standard form of the Einstein-Hilbert Lagrangian of GR. In particular, we can take $\mathcal{L%
}=R$, on a pseudo--Riemannian manifold, $V,$\ where $R$ is the Ricci scalar
associated with the Levi--Civita connection $\nabla $.  Other corresponding functionals depend
on the torsion tensor, $T_{\beta \gamma }^{\alpha },$ the matter energy--momentum tenso,  $T_{\beta \gamma },$ and/or trace $T=T_{\alpha }^{\alpha },$ on the
Ricci scalar $\mathbf{R}$ for a generalized connection $\mathbf{D}$ etc.
The physical/ geometric objects for such theories are defined on the tangent
bundle $TV,$ or on a generalized nonholonomic manifold $\mathbf{V}$ with a
prescribed nonholonomic structure of frames, distortions of geometric
objects etc \cite{revfmod}. This includes theories with Lorentz violations,
nonlinear dispersion relations,  locally anisotropic re-scaling and
effective polarizations of constants which provides a deeper understanding of the
relations among $f$--modified, Ho\v{r}ava--Lifshitz and Finsler-like
theories.

For MGTs with nontrivial torsion, we can naturally find motivations for the presence of
fermionic (gravitino) terms. Such configurations are generically present in
SG theories via four--gravitino self--interaction terms which may produce condensates of the gravitino field under well--defined conditions.
For certain models, the gravitino may acquire mass but leave the graviton
massless. In general, we can consider bi-metric and massive theories \cite%
{massgr} for off--diagonal configurations encoding MGTs contributions \cite%
{voffdmgt}. In the model \cite{jas}, the mass of the gravitino is generated
dynamically to be of order of the Planck scale. It was conjectured that the
dynamical breaking of local supersymmetry is a result of the formation of
condensates of the gravitino filed. The approach was studied for the Deser-Zumino super--Higgs effect 
\cite{deser} which resulted from a coupling of the supergravity action with the nonlinear
Volkov--Akulov action\cite{volkov} and the Goldsone particle. Assuming a
F--type spontaneous global supersymmetry breaking, the final result is a
Majorana spin 1/2 fermion for the so--called Godstino field.

The authors of \cite{buch} criticised the flat Minkowski spacetime approach
\cite{jas} because it ignored the quantum fluctuations of the metric field.
Such fluctuations over the metric backgrounds introduce imaginary parts to the
one--loop effective potential in four--dimensional supergravity theories for
any non--trivial value of the gravitino condensate field.\footnote{%
Imaginary parts may result in the instability of the
non--trivial--gravitino-condensate (breaking $\mathcal{N}=1$ supergravity).}
In general, this does not depend on the introduction of a possible
background cosmological constant $\Lambda $ which may replace \cite{fradkin}
the ultra violet cut--off used in the flat spacetime. Recently, the
arguments of \cite{buch} were reconsidered in \cite{mavr1} by incorporating the
super--Higgs effect with the aim to build a simple model of dynamical
breaking of local supersymmetry by means of the gravitino--torsion
self--interactions. Torsion fields exist in various (super) string
and gravity theories and it is naturally to consider that they may be
responsible for the local supersymmetry breaking.  In another turn, various models of MGTs were
studied recently for certain functional modifications of the Lagrangians with
anisotropic symmetries on certain effective pseudo--Riemannian spaces (which
in terms of the Levi--Civita (LC) connection are torsionless).

Our goal is to study three problems based in conjecturing that a dynamical
breaking of supergravity is possible via the formation of condensates of the
gravitino fields \ when the spacetime is not flat in GR or in MGT. The problems are : 1) \ Will
super--Higgs generalized effects help in the construction of modified supergravity
theories (MSGT)? 2) How such super--Higgs models are related to the various
classes of generic off--diagonal solutions that can be constructed in GR
and MGTs?\ 3) How a mechanism of dynamical SG--breaking can be related to
an accelerating universe and to dark energy and dark matter cosmology?  In this
article, we shall develop the approaches taken from \cite{buch,mavr1} to
investigate possible dynamical breaking scenarios in supergravity which may
be connected with certain limits (not only of GR) but of MGTs and to study the
corresponding physically important classes of solutions which can be
realized as effective off--diagonal Einstein manifolds.

Let us provide certain motivations for our work. The main idea comes from a
(very surprising) decoupling property of the vacuum and non--vacuum fundamental
field equations in certain classes of gravity theories. In a simple way,
such a property can be proven if one works with an "auxiliary" connection when
certain associated systems of nonlinear PDE decouple with respect to certain
classes of nonholonomic frames. Hence, we can integrate such systems in
very general forms which allows us to generate and study new classes of generic
off--diagonal solutions.\footnote{%
The corresponding metrics cannot be diagonalized by coordinate transformations.}
Usually, an auxiliary connection has a nontrivial nonholonomically induced
torsion which in GR is completely determined by certain generic
off--diagonal terms in the metrics and/or after imposing certain classes of nonholonomic
(equivalently, anholonomic, i.e. non--integrable) constraints on the
gravitational and matter fields dynamics.  We note that, in general, a
nonholonomically induced torsion is different from the torsion fields in the
Einstein--Cartan gauge theories of gravity, or in string gravity theory. In our
approach, the torsion is induced by the off--diagonal terms of the metric field and
(not obligatory) subjected to additional algebraic or dynamical field
equations. The geometric formalism with nonholonomic configurations can be
extended to various classes of commutative and noncommutative, or
supersymmetric gravity and field theories, by relating nonholonomic frame
deformations to nontrivial torsion fields of a different nature. Having
constructed certain general integral varieties of solutions, we can impose
almost always certain classes of constraints when the auxiliary connection
transforms into the Levi-Civita (LC) one. Such zero--torsion constraints can be imposed
additionally after certain classes of generalized solutions are found, i.e.
generic off--diagonal solutions for nonlinear systems can be restricted to have
torsionless configurations if certain nontrivial solutions were already
found for the decoupled configurations. The end result is that it was possible to construct
a geometric formalism, the so--called anholonomic frame deformation method
(AFDM), for constructing exact solutions in gravity \cite{voffds}.

Our goal is to demonstrate in detailed that a nonholonomic (and alternative
off--diagonal) incorporation of the super--Higgs effect may result in the
dynamical breaking of MSGTs, see examples of such works and applications in
cosmology \cite{ketov}. This can be realized prior to the coupling to matter and
gauge fields. The AFDM allows us to encode various modifications of
(super) gravitational theories into off--diagonal Einstein spaces whose
effective cosmological constant which can be chosen to be positive for
certain well--defined conditions. We can perform a one--loop effective
potential calculation considering the metric fluctuations fully and (not small)
off--diagonal deformations, about the de Sitter backgrounds. Within the framework of the
correspondingly defined nonholonomic frames, any (weak) quantum
gravitational effects reveals the existence of non--trivial vacua with no
imaginary parts. In this way, we cured the problems mentioned in \cite{buch} and
generalized the constructions of \cite{mavr1} for various classes of MGTs.
The effective positive cosmological constant (which in the AFDM is a simple
parameter used for the re--definition of an integrating function) acquires a
physical meaning and becomes responsible for the vanishing of the effective
vacuum energy. This is due to the super--Higgs effect which is determined by the one-loop
low--energy effective action of supergravity.

The article is organized as follows. In Sec. \ref{s2}, we survey a geometric
technique for generating off--diagonal solutions with an associated torsion
structure in MGT and GR.  Explicit examples are considered of three-dimensional
gravitational solitonic waves possessing different properties and having possible effects in supergravity and modified gravity. Sec. %
\ref{s3} is devoted to $\mathcal{N}=1,d=4$ modified supergravity models with
nonholonomic super--Higgs effects and goldstino coupling. In Sec. \ref{s4},
we perform a one--loop calculation in order to compute the partition functions
in MGT deformations with nonholonomic bosonic and fermionic configurations.
We discuss in brief the stability conditions determined by the nonholonomic
constraints and nonlinear off--diagonal interactions and also discuss how such issues can
be related to solitonic solutions in gravity theories.\ Finally, a
discussion of the results and the conclusions are presented in Sec. \ref{s5}.

\section{Encoding MGTs into Off--Diagonal Metrics and Torsions}

\label{s2}

Various classes of MGTs were elaborated with the aim to explain
observational data in modern cosmology and solve the dark energy and dark
matter problems. In this section we outline a geometric approach which
allows us to encode various classes of solutions of field equations in MGTs
into the corresponding off--diagonal metrics that define the (effective) Einstein
manifolds. We summarize also the anholonomic frame deformation method, AFDM,
for constructing off--diagonal solutions in gravity and Ricci flow \
theories \cite{voffds,vrevlf}. This is possible for metric-compatible linear
connections with noholonomically induced torsion and that are completely defined by the
metric tensors and the anholonomy frame coefficients.  After imposing additional
nonintregrable constraints, we can "extract" solutions for the Levi--Civita
connection $\nabla .$

\subsection{Preliminaries}

Our key geometric idea is that GR and various MGTs can be described
equivalently in terms of the geometric data $(\mathbf{g},\nabla )$ and/ or $(%
\mathbf{g},\widehat{\mathbf{D}})$ using an "auxiliary" canonical
distinguished connection $\widehat{\mathbf{D}}$ which also is completely
defined by the same metric structure $\mathbf{g}$. In general, $\widehat{%
\mathbf{D}}$ has a nontrivial nonholonomically induced torsion $\widehat{%
\mathcal{T}}$ determined by certain generic off--diagonal coefficients of $%
\mathbf{g}$.\footnote{%
Such a metric is considered generic off--diagonal if it cannot be
diagonalized via coordinate transformations.} Using $\widehat{\mathcal{T}}$ and
the corresponding classes of exact solutions in MGT, we find additional
mechanisms for dynamical supersymmetry breaking if the constructions are
extended to SG.

\subsubsection{N--adapted frames, d--metrics and d--connections}

Any metric structure $\mathbf{g}$ on a pseudo--Riemannian manifold $\mathbf{%
V,}$ for instance, of signature $\left( +,+,+,-\right) ,$ can be written in
two equivalent forms:

\begin{enumerate}
\item for a local coordinate co--base,
\begin{equation*}
\mathbf{g}=\underline{g}_{\alpha \beta }du^{\alpha }\otimes du^{\beta },
\end{equation*}%
with coefficients
\begin{equation}
\underline{g}_{\alpha \beta }=\left[
\begin{array}{cc}
g_{ij}+N_{i}^{a}N_{j}^{b}g_{ab} & N_{j}^{e}g_{ae} \\
N_{i}^{e}g_{be} & g_{ab}%
\end{array}%
\right] ,  \label{ansatz}
\end{equation}%
for coordinates $u=(x,y),$ or $u^{\alpha }=(x^{i},y^{a}),$ and $h$--indices $%
i,j,...=1,2,...n$ and $v$--indices $a,b,...=n+1,n+2,...,n+m;$ when $\dim
\mathbf{V}=n+m,$ for $n,m\geq 2$.\footnote{The Einstein summation rule on the \textquotedblright
up-low\textquotedblright\ cross indices is used unless otherwise stated. Boldface letters shall be used in order to emphasize that a
N--connection splitting has been considered on the spacetime manifold $\mathbf{V}=(V,\mathbf{N)}$.
In order to avoid cumbersome formulas, we shall write off--diagonal terms in explicit form only for 4-d configurations (if this does not result in ambiguities). In general, our approach can be elaborated for arbitrary finite dimensions.}

\item We can represent $\mathbf{g}$ as a distinguished metric (in brief,
\textit{d--metric),}\footnote{this term is largely used in some monographs on differential geometry but also in some  series of works on gravity and mathematical particle physics \cite{voffds,vrevlf,vspinor,vacarsolitonhier}, see also references therein; it was introduced with three main purposes:\ to elaborate geometric constructions adapted with respect to certain natural/ conventional h- and v--splitting which allows to consider further parameterizations in order to decouple gravitational field equations for corresponding classes of nonlinear and linear connections and, finally, to perform deformation quantization of such nonlinear systems}
\begin{equation}
\mathbf{g}=g_{\alpha }(u)\mathbf{e}^{\alpha }\otimes \mathbf{e}^{\alpha
}=g_{i}(x^{k})dx^{i}\otimes dx^{i}+g_{a}(x^{k},y^{b})\mathbf{e}^{a}\otimes
\mathbf{e}^{a}.  \label{dm1}
\end{equation}
\end{enumerate}

In the N--adapted form the nonholonomic frames $\mathbf{e}_{\nu }=(\mathbf{e}%
_{i},e_{a})$ and $\mathbf{e}^{\mu }=(e^{i},\mathbf{e}^{a}),$ are
\begin{eqnarray}
\mathbf{e}_{i} &=&\partial /\partial x^{i}-\ N_{i}^{a}(u)\partial /\partial
y^{a},\ e_{a}=\partial _{a}=\partial /\partial y^{a};  \label{dder} \\
e^{i} &=&dx^{i},\ \mathbf{e}^{a}=dy^{a}+\ N_{i}^{a}(u)dx^{i}.  \label{ddif}
\end{eqnarray}%
They satisfy the relations
\begin{equation}
\lbrack \mathbf{e}_{\alpha },\mathbf{e}_{\beta }]=\mathbf{e}_{\alpha }%
\mathbf{e}_{\beta }-\mathbf{e}_{\beta }\mathbf{e}_{\alpha }=W_{\alpha \beta
}^{\gamma }\mathbf{e}_{\gamma },  \label{nonholr}
\end{equation}%
whose anholonomy coefficients are $W_{ia}^{b}=\partial
_{a}N_{i}^{b},W_{ji}^{a}=\Omega _{ij}^{a}=\mathbf{e}_{j}\left(
N_{i}^{a}\right) -\mathbf{e}_{i}(N_{j}^{a}).$ In these formulas, the
coefficients $\Omega _{ij}^{a}$ define the N--connection curvature, see below.

The coefficients $\mathbf{N}=\{N_{i}^{a}(u)\}=N_{i}^{a}(x,y)dx^{i}\otimes
\partial /\partial y^{a}$ define a nonholonomic (equivalently,
non--integrable, or anholonomic) horizontal (h) and vertical (v) splitting,
or a nonlinear connection (\textit{N--connection}) structure if a Whitney
sum $\mathbf{N}$ is obtained for its tangent space $T\mathbf{V}$,
\begin{equation}
\mathbf{N}:\ T\mathbf{V}=hT\mathbf{V}\oplus vT\mathbf{V}.  \label{whitn}
\end{equation}

We say that a linear connection is a distinguished connection,\textit{\
d--connecti\-on,} $\mathbf{D}=(hD,vD),$ if it preserves under parallelism a $%
h$-$v$--splitting (\ref{whitn}). It defines a covariant derivative, $\mathbf{%
D}_{\mathbf{X}}\mathbf{Y}$, for a d--vector field $\mathbf{Y}$ in the
direction of a d--vector $\mathbf{X}.$ In general, any vector field $Y(u)\in
T\mathbf{V}$ can be parameterized as a d--vector, $\mathbf{Y}=$ $\mathbf{Y}%
^{\alpha }\mathbf{e}_{\alpha }=\mathbf{Y}^{i}\mathbf{e}_{i}+\mathbf{Y}%
^{a}e_{a},$ or $\mathbf{Y}=(hY,vY),$ with $hY=\{\mathbf{Y}^{i}\}$ and $vY=\{%
\mathbf{Y}^{a}\}.$ The local coefficients $\mathbf{D}=\{\mathbf{\Gamma }_{\
\alpha \beta }^{\gamma }=(L_{jk}^{i},L_{bk}^{a},C_{jc}^{i},C_{bc}^{a})\}$
are computed as h--v--components of $\mathbf{D}_{\mathbf{e}_{\alpha }}%
\mathbf{e}_{\beta }:=$ $\mathbf{D}_{\alpha }\mathbf{e}_{\beta }$ using $%
\mathbf{X}=\mathbf{e}_{\alpha }$ and $\mathbf{Y}=\mathbf{e}_{\beta }.$ The
terms d--vector, d--tensor etc can be used for any vector, tensor value with
coefficients defined in the N--adapted form with respect to the necessary types of
tensor products of bases (\ref{dder}) and (\ref{ddif}).

For any d--connection structure $\mathbf{D,}$ we can construct three
fundamental geometric objects: the d--torsion, $\mathcal{T}(\mathbf{X,Y}):=%
\mathbf{D}_{\mathbf{X}}\mathbf{Y}-\mathbf{D}_{\mathbf{Y}}\mathbf{X}-[\mathbf{%
X,Y}],$ the nonmetricity, $\mathcal{Q}(\mathbf{X}):=\mathbf{D}_{\mathbf{X}}%
\mathbf{g},$ and the d--curvature, $\mathcal{R}(\mathbf{X,Y}):=\mathbf{D}_{%
\mathbf{X}}\mathbf{D}_{\mathbf{Y}}-\mathbf{D}_{\mathbf{Y}}\mathbf{D}_{%
\mathbf{X}}-\mathbf{D}_{\mathbf{[X,Y]}}.$ Respectively, the N--adapted
coefficients,
\begin{eqnarray*}
&&\mathcal{T}=\{\mathbf{T}_{\ \alpha \beta }^{\gamma }=\left( T_{\
jk}^{i},T_{\ ja}^{i},T_{\ ji}^{a},T_{\ bi}^{a},T_{\ bc}^{a}\right) \},%
\mathcal{Q}=\mathbf{\{Q}_{\ \alpha \beta }^{\gamma }\}, \\
&&\mathcal{R}\mathbf{=}\mathbf{\{R}_{\ \beta \gamma \delta }^{\alpha }%
\mathbf{=}\left( R_{\ hjk}^{i}\mathbf{,}R_{\ bjk}^{a}\mathbf{,}R_{\ hja}^{i}%
\mathbf{,}R_{\ bja}^{c},R_{\ hba}^{i},R_{\ bea}^{c}\right) \},
\end{eqnarray*}
are computed by introducing $\mathbf{X}=\mathbf{e}_{\alpha }$ and $\mathbf{Y}%
=\mathbf{e}_{\beta },$ and $\mathbf{D}=\{\mathbf{\Gamma }_{\ \alpha \beta
}^{\gamma }\}$ into above formulas, see details in Refs. \cite{voffds}.

A d--connection $\mathbf{D}$ is compatible with a d--metric $\mathbf{g}$ if
and only if $\mathcal{Q}=\mathbf{Dg}=0.$ The Levi--Civita (LC) connection, $%
\nabla ,$ is the unique metric compatible and torsionless linear connection
determined by a metric structure $\mathbf{g.}$ We can always consider a
N--adapted distortion of a d--connection $\mathbf{D}$ when
\begin{equation}
\mathbf{D}=\nabla +\mathbf{Z}.  \label{distr}
\end{equation}%
The N--adapted coefficients are labeled respectively: $\nabla =\{\Gamma _{\
\beta \gamma }^{\alpha }\}$ and, for the distortion d--tensor, $\mathbf{Z}=\{%
\mathbf{Z}_{\ \beta \gamma }^{\alpha }\}.$

\subsubsection{Nonholonomic torsions in (pseudo) Riemannian geometry}

Prescribing a N--connection $\mathbf{N,}$ we can work equivalently with $%
\nabla $ and, for instance, the so--called canonical d--connection, $%
\widehat{\mathbf{D}},$ when
\begin{equation}
(\mathbf{g,N})\rightarrow \{
\begin{array}{ccc}
\mathbf{\nabla :} & \mathbf{\nabla g}=0;\ ^{\nabla }\mathcal{T}=0; &  \\
\widehat{\mathbf{D}}: & \widehat{\mathbf{D}}\mathbf{g}=0;\ h\widehat{\mathcal{T}}=0,v\widehat{\mathcal{T}}=0,hv\widehat{\mathcal{T}}\neq 0; &
\end{array}
\label{cand}
\end{equation}%
and all geometric objects are completely defined by the same metric
structure. In above formulas,   $h\widehat{\mathcal{T}}$ and $v\widehat{\mathcal{T}}$ are torsion components on, respectively, h- and v--subspace when a nontrivial $hv\widehat{\mathcal{T}}$ is possible for horizontal-vertical components because of nonholonomic structure.

The Ricci tensors of $\widehat{\mathbf{D}}$ and $\nabla $ are computed in the
standard form, $\ \widehat{\mathcal{R}}ic=\{\widehat{\mathbf{R}}_{\ \alpha \beta
 }:=\widehat{\mathbf{R}}_{\ \alpha \beta \gamma }^{\gamma }\}$ and $%
Ric=\{R_{\ \alpha \beta }:=R_{\ \alpha \beta \gamma }^{\gamma }\}.$ The
Ricci d--tensor $\widehat{\mathcal{R}}ic$ is characterized by N--adapted
coefficients
\begin{equation}
\widehat{\mathbf{R}}_{\alpha \beta }=\{\widehat{R}_{ij}:=\widehat{R}_{\
ijk}^{k},\ \widehat{R}_{ia}:=-\widehat{R}_{\ ika}^{k},\ \widehat{R}_{ai}:=%
\widehat{R}_{\ aib}^{b},\ \widehat{R}_{ab}:=\widehat{R}_{\ abc}^{c}\},
\label{driccic}
\end{equation}
and scalar curvature,
\begin{equation}
\ \widehat{\mathbf{R}}:=\mathbf{g}^{\alpha \beta }\widehat{\mathbf{R}}%
_{\alpha \beta }=g^{ij}\widehat{R}_{ij}+g^{ab}\widehat{R}_{ab},
\label{sdcurv}
\end{equation}%
which is alternative to the LC--scalar curvature, $\ R:=\mathbf{g}^{\alpha
\beta }R_{\alpha \beta }.$

Any (pseudo) Riemannian geometry can be equivalently described by both
geometric data $\left( \mathbf{g,\nabla }\right) $ and/or $(\mathbf{g,N,}%
\widehat{\mathbf{D}})$ and canonical distortion relations, $\widehat{%
\mathcal{R}}=\ ^{\nabla }\mathcal{R+}\ ^{\nabla }\mathcal{Z}$ and $\widehat{%
\mathcal{R}}ic=Ric+\widehat{\mathcal{Z}}ic.$ Nevertheless, gravitational and
matter field theories for different $\mathbf{\nabla }$ and $\widehat{\mathbf{%
D}}$ are not equivalent if certain additional conditions are not imposed.
The Einstein d--tensor of $\widehat{\mathbf{D}},$
\begin{equation}
\widehat{\mathbf{E}}_{\alpha \beta }:=\widehat{\mathbf{R}}_{\alpha \beta }-%
\frac{1}{2}\mathbf{g}_{\alpha \beta }\ \widehat{\mathbf{R}},  \label{enstdt}
\end{equation}%
and energy--momentum d--tensors, $\widehat{\mathbf{T}}_{\alpha \beta }:=-%
\frac{2}{\sqrt{|\mathbf{g}_{\mu \nu }|}}\frac{\delta (\sqrt{|\mathbf{g}_{\mu
\nu }|}\ \ ^{m}\widehat{\mathcal{L}})}{\delta \mathbf{g}^{\alpha \beta }}$
constructed for a Lagrange density $\ ^{m}\mathcal{L}$, are different from
those used for $\nabla .$ For a nontrivial N--connection structure, we
construct a nonholonomic deformation of the Einstein gravity, with $\nabla
\rightarrow $ $\widehat{\mathbf{D}}=\nabla +\widehat{\mathbf{Z}},$ with
gravitational field equations
\begin{equation}
\widehat{\mathbf{R}}_{\alpha \beta }=\kappa ^{2}(\widehat{\mathbf{T}}%
_{\alpha \beta }-\frac{1}{2}\mathbf{g}_{\alpha \beta }\widehat{\mathbf{T}}),
\label{nheeq}
\end{equation}%
for a conventional gravitational constant $\kappa ^{2}$ and $\widehat{%
\mathbf{T}}:=\mathbf{g}^{\mu \nu }\widehat{\mathbf{T}}_{\mu \nu }.$

The equations (\ref{nheeq}) contain nontrivial N--adapted coefficients for the
torsion, {\small
\begin{equation}
\widehat{T}_{\ jk}^{i}=\widehat{L}_{jk}^{i}-\widehat{L}_{kj}^{i},\widehat{T}%
_{\ ja}^{i}=\widehat{C}_{jb}^{i},\widehat{T}_{\ ji}^{a}=-\Omega _{\ ji}^{a},%
\widehat{T}_{aj}^{c}=\widehat{L}_{aj}^{c}-e_{a}(N_{j}^{c}),\widehat{T}_{\
bc}^{a}=\ \widehat{C}_{bc}^{a}-\ \widehat{C}_{cb}^{a}.  \label{dtors}
\end{equation}%
} which are determined completely by the coefficients $\mathbf{g}_{\alpha \beta
} $ (\ref{dm1}), $N_{i}^{a}$ and their derivatives. Levi-Civita (LC)--configurations can
be extracted from certain classes of solutions of the equations (\ref{nheeq}) if
additional conditions are imposed resulting in zero values for the
canonical d--torsion, $\widehat{\mathcal{T}}=0$.

\subsection{Equivalent modelling of MGTs}

Let us state the conditions when the three classes of MGTs can be equivalently
modelled by the solutions of certain corresponding field equations. We consider such
actions and functionals for gravitational, $\ ^{g}S,$ and matter, $\ ^{m}S,$
fields:{\small
\begin{eqnarray}
\mathcal{S} &=&\ ^{g}\mathcal{S}+\ ^{m}\mathcal{S}=\frac{1}{2\kappa ^{2}}%
\int f(R)\sqrt{|g|}d^{4}u+\int \ ^{m}\mathcal{L}\sqrt{|g|}d^{4}u  \label{act}
\\
&=&\ ^{g}\widehat{\mathbf{S}}+\ ^{m}\widehat{\mathbf{S}}=\frac{1}{2\kappa
^{2}}\int \widehat{\mathbf{f}}(\widehat{\mathbf{R}})\sqrt{|\mathbf{g}|}%
\mathbf{d}^{4}u-\frac{\mathring{\mu}^{2}}{4}\mathcal{U}(\mathbf{g}_{\mu \nu
},\mathcal{K}_{\alpha \beta })+\int \ ^{m}\widehat{\mathbf{L}}\sqrt{|\mathbf{%
g}|}\mathbf{d}^{4}u  \notag \\
&=&\ ^{g}\mathbf{\check{S}}+\ ^{m}\mathbf{\check{S}}=\frac{1}{2\kappa ^{2}}%
\int \mathbf{\check{R}}\sqrt{|\mathbf{\check{g}}|}\mathbf{d}^{4}u+\check{\Lambda}\int \sqrt{|\mathbf{\check{g}}|}\mathbf{d}^{4}u,  \label{mgts}
\end{eqnarray}%
} defined on a pseudo--Riemannian manifold $\mathbf{V}$ by the same metric
field $\mathbf{g}=\{g_{\mu \nu }\},$ but up to different classes of
nonholonomic frame transformations and deformations with $(g,\nabla )\sim (%
\mathbf{g},\mathbf{N},\widehat{\mathbf{D}})\sim (\mathbf{\check{g},\check{N},%
\check{D}}).$ The equality symbols specify that we shall work with one action written in three different forms when the matter $\ ^{m}\mathbf{\check{S}}$ is encoded into and effective term with $\check{\Lambda}$. We shall explain below the nature of the all terms in above actions.

The theories with actions of type (\ref{act}) generalize the so--called
modified $f(R)$ gravity, see reviews and original results in \cite{revfmod},
and the ghost--free massive gravity (by de Rham, Gabadadze and Tolley, dRGT)
\cite{drg}. We use the natural units of  $\hbar =c=1$ and the square of the Planck
mass is $M_{Pl}^{2}=1/8\pi G=\kappa ^{-2}$ with the 4--d Newton constant $G.$
It is written $\mathbf{d}^{4}u$ in bold face instead of $d^{4}u$ because the
N--elongated differentials are used (\ref{ddif}) and $\mathring{\mu}=const$ is taken for
the mass of the graviton. We can fix conditions of the type
\begin{equation}
\widehat{\mathbf{f}}(\widehat{\mathbf{R}})-\frac{\mathring{\mu}^{2}}{4}%
\mathcal{U}(\mathbf{g}_{\mu \nu },\mathcal{K}_{\alpha \beta })=\mathbf{f}(%
\widehat{\mathbf{R}}),\mbox{ \ or \ }\widehat{\mathbf{f}}(\widehat{\mathbf{R}%
})=f(R),\mbox{ \ or \ }\widehat{\mathbf{f}}(\widehat{\mathbf{R}})=R,
\label{mgrfunct}
\end{equation}%
which allows us to extract LC--configurations depending on the type of
models and classes of solutions we are going to study.  To analyze possible
scenarios of dynamical supersymmetry breaking it is more convenient to work
with geometric \ variables and \ field equations determined by $\widehat{%
\mathbf{D}}$ and $\widehat{\mathbf{R}}.$

The equations of motion for massive MGT (\ref{act}) can be written in the
variables $(\mathbf{g},\mathbf{N},\widehat{\mathbf{D}}),$
\begin{equation}
(\partial _{\widehat{\mathbf{R}}}\widehat{\mathbf{f}})\widehat{\mathbf{R}}%
_{\mu \nu }-\frac{1}{2}\widehat{\mathbf{f}}(\widehat{\mathbf{R}})\mathbf{g}%
_{\mu \nu }+\mathring{\mu}^{2}\mathbf{X}_{\mu \nu }=\kappa ^{2}\widehat{%
\mathbf{T}}_{\mu \nu },  \label{mgrfe}
\end{equation}%
where $M_{Pl}$ is the Plank mass. For simplicity, we shall consider matter
actions which depend only on the coefficients of a metric field and not on its
derivatives,
\begin{equation*}
\widehat{\mathbf{T}}^{\alpha \beta }=\ ^{m}\widehat{\mathbf{L}}\ \mathbf{g}%
^{\alpha \beta }+2\delta (\ ^{m}\widehat{\mathbf{L}})/\delta \mathbf{g}%
_{\alpha \beta },
\end{equation*}%
see details of the variational methods in \cite{drg}; we shall follow some
conventions from \cite{kobayashi}). \ We are able to construct solutions in
MGT in explicit form if we fix the coefficients $\{N_{i}^{a}\}$ of $\mathbf{N%
}$ and local frames for $\widehat{\mathbf{D}}$ when $\widehat{\mathbf{R}}%
=const$ and $\partial _{\alpha }\widehat{\mathbf{f}}(\widehat{\mathbf{R}}%
)=(\partial _{\widehat{\mathbf{R}}}\widehat{\mathbf{f}})\times \partial
_{\alpha }\widehat{\mathbf{R}}=0$ but, in general, $\partial _{\alpha
}f(R)\neq 0.$\ For $\widehat{\mathbf{D}}\rightarrow \nabla ,$ we obtain $%
\widehat{\mathbf{R}}_{\mu \nu }\rightarrow R_{\mu \nu }$ with a standard
Ricci tensor $R_{\mu \nu }$ for $\nabla .$

The effective energy--momentum tensor $\mathbf{X}_{\mu \nu }$ in (\ref{mgrfe}%
) is defined by the potential of the graviton $\mathcal{U}=\mathcal{U}%
_{2}+\alpha _{3}\mathcal{U}_{3}+\alpha _{4}\mathcal{U}_{4},$ where $\alpha
_{3}$ and $\alpha _{4}$ are free parameters. The values $\mathcal{U}_{2},%
\mathcal{U}_{3}$ and $\mathcal{U}_{4}$ are certain polynomials of traces of
some other polynomials of the matrix $\mathcal{K}_{\mu }^{\nu }=\delta _{\mu
}^{\nu }-\left( \sqrt{g^{-1}\Sigma }\right) _{\mu }^{\nu }$ for a tensor
determined by four St\"{u}ckelberg fields $\phi ^{\underline{\mu }}$ as
\begin{equation}
\Sigma _{\mu \nu }=(\mathbf{e}_{\mu }\phi ^{\underline{\mu }})(\mathbf{e}%
_{\nu }\phi ^{\underline{\nu }})\eta _{\underline{\mu }\underline{\nu }},
\label{bm}
\end{equation}%
when $\eta _{\underline{\mu }\underline{\nu }}=(1,1,1,-1).$ Following a
series of arguments presented in \cite{kobayashi}, when the parameters are chosen
$\alpha _{3}=(\alpha -1)/3,\alpha _{4}=(\alpha ^{2}-\alpha +1)/12$ and which is useful
for avoiding potential ghost instabilities, we can fix
\begin{equation}
\mathbf{X}_{\mu \nu }=\alpha ^{-1}\mathbf{g}_{\mu \nu }  \label{cosmconst}
\end{equation}%
with a constant $\alpha $ which induces an additional term for an (effective)
cosmological \ constant. For such parameterizations, the field equations in
MGT can be written in an effective form which is very similar to (\ref{nheeq}%
),
\begin{equation}
\widehat{\mathbf{R}}_{\mu \nu }=\widehat{\mathcal{Y}}_{\alpha \beta }
\label{mgrfec}
\end{equation}%
with effective source%
\begin{equation}
\widehat{\mathcal{Y}}_{\alpha \beta }=(\partial _{\widehat{\mathbf{R}}}%
\widehat{\mathbf{f}})^{-1}[(\frac{1}{2}\widehat{\mathbf{f}}-\mathring{\mu}%
^{2})\mathbf{g}_{\alpha \beta }+\kappa ^{2}\widehat{\mathbf{T}}_{\alpha
\beta }].  \label{dsourc}
\end{equation}%
Redefining the generating function and effective source (see below formula
(\ref{aux2}) and related details) we can encode the MGT contributions into
generic off--diagonal terms of an effective d--metric $\mathbf{\check{g}}$ (%
\ref{dm1}) and canonical d--connection $\mathbf{\check{D}}$ as solutions of the
nonholonomically deformed Einstein manifolds with field equations determined
by the action (\ref{mgts}),%
\begin{equation}
\mathbf{\check{R}}_{\mu \nu }=\check{\Lambda}\mathbf{\check{g}}_{\alpha
\beta }.  \label{efes}
\end{equation}

In the next subsection, we shall explain how the equations and solutions of (\ref{mgrfec}) can be
transformed into similar data as that described by eq-(\ref{efes}) and conversely.

\subsection{ The AFDM in MGT}

We summarize the anholonomic frame method, AFDM, which allows us to
construct generic off--diagonal solutions in MGTs \cite{voffdmgt,voffds}.

\subsubsection{Decoupling property for the canonical d--connection with
respect to N--adapted frames}

We can consider such matter fields and N--adapted frame and coordinate
transformations when a source (\ref{dsourc}) is parametrized as
\begin{eqnarray}
\widehat{\mathcal{Y}}_{\alpha \beta } &\rightarrow &diag[\Upsilon
_{1}=\Upsilon _{2},\Upsilon _{2}=\ ^{h}\Upsilon (x^{i}),\Upsilon
_{3}=\Upsilon _{4},\Upsilon _{4}=\ ^{v}\Upsilon (x^{i},y^{4})],
\label{dsours1} \\
\mbox{ or } &\rightarrow &{\Lambda }\ \mathbf{g}_{\alpha \beta },\ {\Lambda }%
=const.  \label{dsours2}
\end{eqnarray}
For simplicity, we can consider effective matter sources and "prime" metrics
with a Killing symmetry in $\partial /\partial _{3},$ i.e. when the effective
matter sources and d--metrics do not depend on the coordinate $y^{3}.$\footnote{%
Solutions with dependencies on $y^{3}$ and non--Killing configurations are
studied in \cite{voffds}.} In brief, the partial derivatives $\partial
_{\alpha }=\partial /\partial u^{\alpha }$ on a 4--d manifold \ will be
written $s^{\bullet }=\partial s/\partial x^{1},s^{\prime }=\partial
s/\partial x^{2},s^{\ast }=\partial s/\partial y^{3},s^{\diamond }=\partial
s/\partial y^{4}.$

The nontrivial components of the Ricci d--tensor (\ref{driccic}) for a
d--metric ansatz (\ref{dm1}) with coefficients
\begin{equation}
\mathbf{g}_{ij}=diag[g_{i}=\epsilon _{i}e^{\psi (x^{k})}],g_{ab}=diag[\omega
^{2}(u^{\alpha })h_{a}(x^{k},y^{4})],  \label{ans2}
\end{equation}%
$\epsilon _{i}=\pm 1$ depending on the signature, with
\begin{equation}
\mathbf{e}_{i}\omega =\partial _{i}\omega -n_{i}\ \omega ^{\ast
}-w_{i}\omega ^{\diamond }=0,  \label{conf}
\end{equation}%
for $N_{i}^{3}=n_{i}(x^{k})$ and $N_{i}^{4}=w_{i}(x^{k},y^{4}),$ are{\small
\begin{eqnarray}
\widehat{R}_{1}^{1} &=&\widehat{R}_{2}^{2}=\frac{1}{2g_{1}g_{2}}%
[-g_{2}^{\bullet \bullet }+\frac{g_{1}^{\bullet }g_{2}^{\bullet }}{2g_{1}}+%
\frac{\left( g_{2}^{\bullet }\right) ^{2}}{2g_{2}}-g_{1}^{\prime \prime }+%
\frac{g_{1}^{\prime }g_{2}^{\prime }}{2g_{2}}+\frac{(g_{1}^{\prime })^{2}}{%
2g_{1}}],  \label{riccidk} \\
\widehat{R}_{3}^{3} &=&\widehat{R}_{4}^{4}=\frac{1}{2h_{3}h_{4}}%
[-h_{3}^{\diamond \diamond }+\frac{\left( h_{3}^{\diamond }\right) ^{2}}{%
2h_{3}}+\frac{h_{3}^{\diamond }h_{4}^{\diamond }}{2h_{4}}],\widehat{R}_{3k}=%
\frac{h_{3}}{2h_{4}}n_{k}^{\diamond \diamond }+(\frac{h_{3}}{h_{4}}%
h_{4}^{\diamond }-\frac{3}{2}h_{3}^{\diamond })\frac{n_{k}^{\diamond }}{%
2h_{4}},  \notag \\
\widehat{R}_{4k} &=&\frac{w_{k}}{2h_{3}}[h_{3}^{\diamond \diamond }-\frac{%
\left( h_{3}^{\diamond }\right) ^{2}}{2h_{3}}-\frac{h_{3}^{\diamond
}h_{4}^{\diamond }}{2h_{4}}]+\frac{h_{3}^{\diamond }}{4h_{3}}(\frac{\partial
_{k}h_{3}}{h_{3}}+\frac{\partial _{k}h_{4}}{h_{4}})-\frac{\partial
_{k}h_{3}^{\diamond }}{2h_{3}}.  \notag
\end{eqnarray}%
}The d--torsion (\ref{dtors}) vanishes if  the
Levi--Civita (LC) conditions are satisfied : $\widehat{L}_{aj}^{c}=e_{a}(N_{j}^{c}),%
\widehat{C}_{jb}^{i}=0,\Omega _{\ ji}^{a}=0.$ Given the ansatz in (\ref{ans2}), we
can write these equations in the form
\begin{eqnarray}
w_{i}^{\diamond } &=&(\partial _{i}-w_{i}\partial _{4})\ln \sqrt{|h_{4}|}%
,(\partial _{i}-w_{i}\partial _{4})\ln \sqrt{|h_{3}|}=0,  \label{lccondb} \\
\partial _{k}w_{i} &=&\partial _{i}w_{k},n_{i}^{\diamond }=0,\partial
_{i}n_{k}=\partial _{k}n_{i}.  \notag
\end{eqnarray}

Using the parametrizations (\ref{dsours1})--(\ref{conf}) for $%
h_{a}^{\diamond }\neq 0$ and $\ ^{h}\Upsilon ,\ ^{v}\Upsilon \neq 0,$ the
MGT fied equations (\ref{mgrfec}) can be written in the form
\begin{eqnarray}
\epsilon _{1}\psi ^{\bullet \bullet }+\epsilon _{2}\psi ^{\prime \prime }
&=&2~^{h}\Upsilon  \label{eq1m} \\
\phi ^{\diamond }h_{3}^{\diamond } &=&2h_{3}h_{4}~^{v}\Upsilon  \label{eq2m}
\\
n_{i}^{\diamond \diamond }+\gamma n_{i}^{\diamond } &=&0,  \label{eq3m} \\
\beta w_{i}-\alpha _{i} &=&0,  \label{eq4m} \\
\partial _{i}\omega -(\partial _{i}\phi /\phi ^{\diamond })\omega ^{\diamond
} &=&0.  \label{confeq}
\end{eqnarray}%
In (\ref{eq3m}) and (\ref{eq4m}) the coefficients are defined as
\begin{equation}
\alpha _{i}=h_{3}^{\diamond }\partial _{i}\phi ,\beta =h_{3}^{\diamond }\
\phi ^{\diamond },\gamma =\left( \ln |h_{3}|^{3/2}/|h_{4}|\right) ^{\diamond
},  \label{abc}
\end{equation}%
where
\begin{equation}
{\phi =\ln |h_{3}^{\diamond }/\sqrt{|h_{3}h_{4}|}|,\mbox{ and/ or }}\Phi
:=e^{{\phi }},  \label{genf}
\end{equation}%
can be considered as a generating function (see below).

There is a decoupling property (with respect to the N--adapted frames and for
the canonical d--connection) of the above system of nonlinear PDEs. This can be
exploited in this form: 1) integrating the 2-d Laplace, or d'Alambert,
equation (\ref{eq1m}) we find solutions for the $h$--coefficients of the
d--metric, $g_{i}=\epsilon _{i}e^{\psi (x^{k})},$ and 2) the solutions for
the $v$--coefficients of the d--metric, $h_{a},$ can be found from (\ref{eq2m})
and (\ref{genf}). \ 3) Then the N--connection coefficients $w_{i}$ and $%
n_{i} $ can be computed respectively from (\ref{eq3m}) and (\ref{eq4m}).

\subsubsection{Generating off--diagonal solutions}

Let us explain how the equations (\ref{eq1m})--(\ref{confeq}) can be solved,
respectively for any sources $~^{h}\Upsilon (x^{k})$ and $\ ^{h}\Upsilon
(x^{i})$ and$\ ^{v}\Upsilon (x^{i},y^{4})$ and generating function ${\phi (x}%
^{k},y^{4}{):}$ The 2-d Laplace/ d'Alambert equation (\ref{eq1m}) is
well--known for physicists and can be solved for any $~^{h}\Upsilon (x^{k}).$

We re--write the system (\ref{eq2m}) and (\ref{genf}) in the form
\begin{equation}
h_{3}h_{4}=\phi ^{\diamond }h_{3}^{\diamond }/2~^{v}\Upsilon \mbox{ and }%
|h_{3}h_{4}|=({h_{3}^{\diamond })}^{2}e^{-2\phi }.  \label{aux21}
\end{equation}
Inserting the first equation into the second one, we can express $%
|h_{3}^{\diamond }|=\frac{(e^{2\phi })^{\diamond }}{4|~^{v}\Upsilon |}=\frac{%
\Phi ^{\diamond }\Phi \ }{2|~^{v}\Upsilon |}.$ This can be integrated with respect to $%
y^{4}$ which results in
\begin{equation}
h_{3}=\ ^{0}h_{3}+\frac{\epsilon _{3}\epsilon _{4}}{4}\int dy^4 \frac{(\Phi
^{2})^{\diamond }}{~^{v}\Upsilon },  \label{aux1a}
\end{equation}
where $\ ^{0}h_{3}=\ ^{0}h_{3}(x^{k})$ and $\epsilon _{3},\epsilon _{4}=\pm
1.$ Using again the first equation from (\ref{aux21}), we obtain
\begin{equation}
h_{4}=\frac{1}{2~^{v}\Upsilon }\frac{\Phi ^{\diamond }}{\Phi }\frac{%
h_{3}^{\diamond }}{h_{3}}.  \label{aux1}
\end{equation}

Having the solutions (\ref{aux1a}) and (\ref{aux1}) for $h_{a}[\Phi ]$, we can
solve equation (\ref{eq3m}) by integrating twice with respect to $dy^{3}$ and
express
\begin{equation}
n_{k}=\ _{1}n_{k}+\ _{2}n_{k}\int dy^4 \ h_{4}/(\sqrt{|h_{3}|})^{3},
\label{n1b}
\end{equation}%
where $\ _{1}n_{k}(x^{i}),\ _{2}n_{k}(x^{i})$ are integration functions. The
rest of the N--connection coefficients can be found from equation (\ref{eq4m}%
) with the coefficients in (\ref{abc}). The solutions of such algebraic linear
equations can be expressed as
\begin{equation}
w_{i}=\frac{\partial _{i}\phi }{\phi ^{\diamond }}=\frac{\partial _{i}\Phi }{%
\Phi ^{\diamond }}=\frac{\partial _{i}(\Phi ^{2})}{(\Phi ^{2})^{\diamond }}.
\label{w1b}
\end{equation}

In general form, the metrics defining the off--diagonal solutions of the MGTs
equations (\ref{mgrfec}) with $\omega $ subjected to the condition (\ref%
{conf}) are parametrized as
\begin{eqnarray}
ds^{2} &=&\epsilon _{i}e^{\psi \lbrack ~^{h}\Upsilon ]}(dx^{i})^{2}+\omega
^{2}(u^{\beta })h_{3}[\Phi ,~^{v}\Upsilon ](dy^{3}+n_{k}[\Phi ,~^{v}\Upsilon
]dx^{k})^{2}  \label{exsol1} \\
&&+\omega ^{2}(u^{\beta })h_{4}[\Phi ,~^{v}\Upsilon ](dy^{4}+w_{k}[\Phi
]dx^{k})^{2},  \notag
\end{eqnarray}%
with the functional form of the coefficients depending on the generating functions $\omega $
and $\Phi ,$  the sources $^{h}\Upsilon $ and $^{v}\Upsilon $ and the integration
functions and parameters as computed above. Such metrics cannot be
diagonalized by coordinate transformations if the anholonomy coefficients $%
W_{\alpha \beta }^{\gamma }$ (\ref{nonholr}) are not trivial. For (\ref%
{exsol1}), the nonholonomically induced torsion (\ref{dtors}) is not zero.
If we fix $\ _{2}n_{k}=0,$ we are able to find $n_{k}=\ _{1}n_{k}(x^{i})$
which have zero torsion limits if the coefficients of the d--metric are chosen
to satisfy, additionally,  the conditions (\ref{lccondb}), see Refs. \cite%
{voffdmgt,voffds} for the explicit geometric \ methods and examples.

\subsubsection{Encoding \ MGTs as nonholonomic \ Einstein manifolds}

There is an important property of the system (\ref{aux21}) and solutions (%
\ref{exsol1}). We can always fix a nontrivial constant $\check{\Lambda}$ and
re--define the generating function, $\Phi \longleftrightarrow \check{\Phi},$
using the formulas
\begin{eqnarray}
\check{\Phi}^{2} &=&\check{\Lambda}^{-1}\left[ \Phi ^{2}|~^{v}\Upsilon
|+\int dy^{4}\ \Phi ^{2}|~^{v}\Upsilon |^{\diamond }\right] ,  \label{aux2}
\\
\Phi ^{2} &=&|\check{\Lambda}||~^{v}\Upsilon |^{-2}\int dy^{4}\ \check{\Phi}%
^{2}|~^{v}\Upsilon |,  \label{aux2b}
\end{eqnarray}%
when $(\Phi ^{2})^{\diamond }/|~^{v}\Upsilon |=(\check{\Phi}^{2})^{\diamond
}/\check{\Lambda}.$ We can simplify the formulas for $h_{a}\rightarrow
\check{h}_{a},$ see (\ref{aux1a}) and (\ref{aux1}), including $\
^{0}h_{3}(x^{k})$ in $\check{\Phi}$ and parametrizing
\begin{equation}
\check{h}_{3}[\check{\Phi}]=\frac{\check{\Phi}^{2}}{4\check{\Lambda}}%
\mbox{
and }\ \check{h}_{4}[\check{\Phi}]=\frac{(\check{\Phi}^{\diamond })^{2}}{%
\check{\Lambda}\Phi ^{2}}=\frac{|\check{\Phi}^{\diamond }~^{v}\Upsilon |^{2}%
}{\check{\Lambda}|\check{\Lambda}|\int dy^{4}\ \check{\Phi}^{2}\
|~^{v}\Upsilon |}.  \label{h34}
\end{equation}%
The conformal factor $\psi \lbrack ~^{h}\Upsilon ]$ with source $%
~^{h}\Upsilon $ is reparametrized by the 2--d coordinate transformations $%
x^{i}\rightarrow \check{x}^{i}(x^{k})$ into $\check{\psi}[\check{\Lambda}],$
when equation (\ref{eq1m}) transforms into
\begin{equation*}
\epsilon _{1}\check{\psi}^{\bullet \bullet }+\epsilon _{2}\check{\psi}%
^{\prime \prime }=2~\check{\Lambda},
\end{equation*}%
where, for instance, $\check{\psi}^{\bullet }=\partial \check{\psi}/\partial
\check{x}^{1}$ and $\check{\psi}^{\prime }=\partial \check{\psi}/\partial
\check{x}^{2}.$

The N--connection coefficients $n_{i}\rightarrow \check{n}_{i}$ and $%
w_{i}\rightarrow \check{w}_{i}$ can be computed as functionals of $\check{%
\Phi}$ and the data $(\check{\Lambda},~^{v}\Upsilon )$ by substituting
respectively $\check{h}_{a}$ (\ref{h34}) and $\Phi ^{2}$ (\ref{aux2b}) into
formulas (\ref{n1b}) and (\ref{w1b}). The equations (\ref{conf}) for the
re--defined conformal factor $\omega \rightarrow \check{\omega}$ transform
into
\begin{equation}
\mathbf{\check{e}}_{i}\check{\omega}=\partial _{i}\check{\omega}-\check{n}%
_{i}\ \check{\omega}^{\ast }-\check{w}_{i}\check{\omega}^{\diamond }=0.
\label{cfeq}
\end{equation}%
We obtain a class of d--metrics
\begin{eqnarray}
ds^{2} &=&\epsilon _{i}e^{\check{\psi}[\check{\Lambda}]}(d\check{x}^{i})^{2}+%
\check{\omega}^{2}(\check{x}^{i},y^{a})\check{h}_{3}[\check{\Phi},\check{%
\Lambda}](dy^{3}+\check{n}_{k}[\check{\Phi},\check{\Lambda},~^{v}\Upsilon ]d%
\check{x}^{k})^{2}  \notag \\
&&+\check{\omega}^{2}(\check{x}^{i},y^{a})\check{h}_{4}[\check{\Phi},\check{%
\Lambda},~^{v}\Upsilon ](dy^{4}+\check{w}_{k}[\check{\Phi},\check{\Lambda}%
,~^{v}\Upsilon ]d\check{x}^{k})^{2},  \label{solut2}
\end{eqnarray}%
defining effective (nonholonomic) Einstein spaces (\ref{efes}) with $\mathbf{%
\check{R}}_{\ \beta }^{\alpha }$ computed with the canonical d--connection $%
\mathbf{\check{D}}.$

The metrics (\ref{solut2}) encode contributions from $f$-modified and/or
massive gravity via $~^{v}\Upsilon $ (\ref{dsours1}). Using re--definitions
of the generating functions and (effective) sources of type (\ref{aux2}) and (%
\ref{aux2b}), we construct nonholonomic deformations of such generic
off--diagonal metrics into the respective classes of solutions described by eq-(\ref{exsol1}) or the
MGT field equations (\ref{mgrfec}).

\subsubsection{Examples: off--diagonal solutions and torsion}

\label{ssectexe}

We analyzed three different classes of off--diagonal solutions (\ref%
{solut2}) encoding MGT configurations characterized by different symmetries
with possible physical implications in classical gravity and dynamical
breaking of supergravity. We proved in \cite{vacarsolitonhier} that (for
certain general conditions) such solutions can be encoded as solitonic
hierarchies and provided various applications in black hole physics, string
and brane gravity with wormholes, flux tubes, nonlinear cosmological
solutions etc. In order to study nonlinear effects, we shall consider three
and two dimensional (3--d and 2--d) solitonic waves with nontrivial
nonholonomically induced torsion of type (\ref{dtors}). If the zero torsion
conditions (\ref{lccondb}) are imposed, we get LC--configurations which may
result in dynamical breaking of local supersymmetry only if additional sources for the
torsion fields are introduced.

Let us parameterize $u^{\alpha }=(x^{i},y^{3},y^{4}=t)$ for a time-like
coordinate $t.$ There are two well known examples when a function $%
q(x^{i},t) $ describes solutions of three-dimensional (3--d) solitonic waves
in terms of  solutions  of

\begin{itemize}
\item the Kadomtsev--Petviashvili, KdP, equation \cite{kadom}
\begin{equation}
q^{\diamond \diamond }\pm (q^{\bullet }+6q\ q^{\prime }+q^{^{\prime \prime
\prime }})^{^{\prime }}=0;  \label{kdp1}
\end{equation}

\item the $2+1$ sine--Gordon equation,%
\begin{equation}
q^{\diamond \diamond }+q^{\prime \prime }+q^{\bullet \bullet }=\sin (q),
\label{sgeq}
\end{equation}%
or other type of solitonic waves and hierarchies which Killing symmetries in
$y^{3},$ or with non--Killing symmetries in the presence of a nontrivial
factor $\check{\omega}^{2}(x^{i},y^{a}).$
\end{itemize}

\paragraph{Generating functions for solitonic waves (for arbitrary sources
from MGTs):}

This class of solutions is defined by $\check{h}=\check{\Phi}^{2}/4\check{%
\Lambda}=q$ in (\ref{h34}). Considering, for simplicity, $\check{\omega}=1$
and substituting $\check{\Phi}=2\sqrt{|\check{\Lambda}q|},$ we generate a
subclass of (\ref{solut2}) of solitonic metrics with Killing symmetry in $%
y^{3}$ {\small
\begin{eqnarray}
ds^{2} &=&\epsilon _{i}e^{\check{\psi}[\check{\Lambda}]}(d\check{x}%
^{i})^{2}+q(dy^{3}+\check{n}_{k}d\check{x}^{k})^{2}+\frac{(q^{\diamond })^{2}%
}{4\check{\Lambda}|\check{\Lambda}|q}\frac{(~^{v}\Upsilon )^{2}}{\int
|q~^{v}\Upsilon |dt}(dt+\frac{\partial _{k}|q|}{|q|^{\diamond }}d\check{x}%
^{k})^{2},  \notag \\
\check{n}_{k} &=&\ _{1}n_{k}(\check{x}^{i})+\ _{2}n_{k}(\check{x}^{i})\int dt%
\frac{(q^{\diamond })^{2}}{|q|^{5/2}}\left( \int_{0}^{t}dt^{\prime
}|q~^{v}\Upsilon (\check{x}^{i},t^{\prime })|\right) ^{-1},
\label{solitonm1}
\end{eqnarray}%
} where the integration functions $\ _{2}n_{k}(\check{x}^{i})$ were
re--defined to include all constants from $h_{4}/(\sqrt{|h_{3}|})^{3}.$ If $%
\ _{2}n_{k}(\check{x}^{i})\neq 0,$ it is not possible to find nonholonomic
distributions resulting in LC--configurations. We need to impose additional
constraints on $\ _{1}n_{k}(\check{x}^{i})$ and source $~^{v}\Upsilon $ in order
to get nonholonomically induced zero torsion configurations. Such special
cases may not result in dynamical supersymmetry breaking.

The solitonic metrics (\ref{solitonm1}) define solitonic solutions with
Killing symmetry in $y^{3}$ for any source $^{v}\Upsilon (\check{x}^{i},t)$
containing information of MGTs. The value $\check{\Lambda}$ is an effective
cosmological constant which can be fixed in order to characterize the
"intensity" of nonlinear interactions. It contributes directly to the bosonic
effective potential (see Section \ref{s4}).

\paragraph{Vertical conformal non--Killing deformations of 3--d solitonic
metrics:}

We can introduce a conformal $v$--factor $\check{\omega}(\check{x}%
^{i},y^{3},t)$ which is a solution of the system of first order PDEs\ (\ref%
{cfeq}),%
\begin{eqnarray}
\check{\omega}^{\bullet }-\check{n}_{1}\check{\omega}^{\ast }-|q|^{\bullet }%
\check{\omega}^{\diamond }/|q|^{\diamond } &=&0,  \label{conf2} \\
\check{\omega}^{^{\prime }}-\check{n}_{2}\check{\omega}^{\ast
}-|q|^{^{\prime }}\check{\omega}^{\diamond }/|q|^{\diamond } &=&0,  \notag
\end{eqnarray}%
where $\check{n}_{k}$ can be taken as in (\ref{solitonm1}). It can be solved
for $\check{\omega}$ for very general classes of solitonic and integrations
functions contained in $q$ and $\check{n}_{k}.$ Conventionally, such metrics
can be parameterized in the form
\begin{eqnarray}
ds^{2} &=&\epsilon _{i}e^{\check{\psi}[\check{\Lambda}]}(d\check{x}^{i})^{2}+%
\check{\omega}^{2}(u^{\alpha })q(dy^{3}+\check{n}_{k}[q,x^{k},^{v}\Upsilon ]d%
\check{x}^{k})^{2}  \notag \\
&&+\check{\omega}^{2}(u^{\alpha })\frac{(q^{\diamond })^{2}}{4\check{\Lambda}%
|\check{\Lambda}|q}\frac{(~^{v}\Upsilon )^{2}}{\int |q~^{v}\Upsilon |dt}(dt+%
\frac{\partial _{k}|q|}{|q|^{\diamond }}d\check{x}^{k}).  \label{solitonm2}
\end{eqnarray}

A nontrivial factor $\check{\omega}$ polarizes the solutions of the nonholonomic
\ Einstein equations with effective cosmological \ constant $\check{\Lambda}%
. $ It contains contributions of MGTs sources $^{v}\Upsilon $ encoded in a
nonlinear wave form in the diagonal and off--diagonal components of the metrics
and with nonholonomically induced torsion.

\paragraph{Off--diagonal "conditional" solitonic configurations:}

\ Both classes of metrics (\ref{solitonm1}) and (\ref{solitonm2}) can be
constructed for arbitrary classes of nonlinear (solitonic) waves which do
not require the condition that the gravitational field equations in a MGT are
transformed, for instance, into a solitonic equation of type (\ref{sgeq}), or (%
\ref{kdp1}). Nevertheless, for certain classes of nonhololonomic
distributions the generalized Einstein equations can be reduced to certain
effective 3--d solitonic equations.

Let us consider the equations (\ref{w1b}) in the form $\Phi ^{\diamond
}w_{i}=\partial _{i}\Phi $. We chose a generating function
\begin{equation}
\Phi =Q(x^{2},t)s(x^{2},t),  \label{2dsolit}
\end{equation}
where $Q=\int dx^{2}(\widetilde{q}^{\diamond \diamond }+[\widetilde{q}%
^{\prime \prime }-\sin (\widetilde{q})]),$ for $\widetilde{q}(x^{2},t).$
This defines a 2--d sine Gordon solution generating off--diagonal \ Einstein
metrics. We get $w_{1}=0$ for $\Phi ^{\bullet }=0$ and
\begin{equation*}
\lbrack Q^{\diamond }s+Qs^{\diamond }]w_{2}=s(\widetilde{q}^{\diamond
\diamond }+[\widetilde{q}^{\prime \prime }-\sin (\widetilde{q})])+Qs^{\prime
}.
\end{equation*}%
If the equation (\ref{sgeq}) is satisfied, $Q=0$ and $w_{2}(x^{2},t)$ is any
function, which allows us to generate vacuum off--diagonal solutions, for $%
\check{\Lambda}=0,$ with two Killing vectors $\partial _{1}$ and $\partial
_{3}.$ Such metrics cannot be represented in a form similar to (\ref%
{solitonm2}). Nevertheless, the nonholonomic induced torsion (\ref{dtors})
may not be zero for such configurations if we do not consider the additional
conditions (\ref{lccondb}) leading to zero torsion. This is an example when the vacuum
Einstein metrics may not break the symmetry.

\paragraph{Off--diagonal "conditional" solitonic configurations with
nontrivial $\check{\protect\omega}:$}

Let us consider a generating function
\begin{equation}
\phi ^{\diamond }=(\ln |\Phi |)^{\diamond }=-2~^{v}\Upsilon (q^{\prime
\prime }+q^{\bullet \bullet }-\sin (q))[(\ln |h_{3}|)^{\diamond }]^{-1}
\label{solitonc}
\end{equation}%
for arbitrary value $h_{3}^{\diamond }\neq 0$ and nonzero $~^{v}\Upsilon $
and $h_{4}=q^{\diamond \diamond }.$ For such configurations, metrics of type
(\ref{exsol1}) with $\omega =1$ are solutions of the MGT field equations (%
\ref{mgrfec}) if $q(x^{i},t)$ is a solution of 3-d sine Gordon equations(\ref%
{sgeq}).

\section{A $\mathcal{N}=1,d=4$ Modified Supergravity Theory}

\label{s3}

Possible nonlinear modifications of gravity are modeled by a nonholonomic
background space $\mathbf{V},$ with N--adapted frames $\mathbf{e}_{\mu }$, d--metric, $\mathbf{g},$ and canonical d--connection, $\widehat{%
\mathbf{D}},$ structures. These structures can be re--defined equivalently in terms of data
of effective Einstein manifolds, $(\mathbf{\check{e}}_{\mu },\mathbf{\check{%
g},\check{D}}),$ described by  solutions of (\ref{efes}). In a simple form, a modified
supergravity theory/ model, MSGT, can be constructed on such nonholonomic
manifolds generalizing the constructions due to Volkov and Akulov \cite{volk}%
.

\subsection{Nonholonomic super-Higgs effects and goldstino coupling}

The Goldstino field $\lambda $ is a four--component Majorana spin - 1/2
spinor. Using N--adapted operators $\mathbf{e}_{\mu },$ the nonholonomic
nonlinear action for the Majorana Goldstino $\lambda ,$ is postulated as
\begin{equation}
\ ^{\lambda }\mathcal{L}=-f^{2}\det \left\vert \delta _{\beta }^{\alpha }+%
\frac{i}{2f^{2}}\overline{\lambda }\gamma ^{\mu }\mathbf{e}_{\mu }\lambda
\right\vert =-f^{2}-\frac{i}{2}\overline{\lambda }\gamma ^{\mu }\mathbf{e}%
_{\mu }\lambda +\ldots ,  \label{actmg}
\end{equation}%
when the determinant is expressed via the weak expansion terms. This defines a
nonholonomic model of $\mathcal{N}=1$ supergravity off--diagonally coupled
to $\lambda .$  Such a Lagrangian arises from a spontaneous/ dynamical breaking
of global supersymmetry when additional nonholonomic
constraints are imposed and the interactions contain certain off--diagonal terms
encoding a MGT model. The gamma matrices $\gamma ^{\mu }$ in the N--adapted
frames are related to the so--called nonholonomic spinor structures and Dirac
operators, see details on nonholonomic supermanifolds and d--spinor geometry
and applications in MGTs in Ref. \cite{vspinor}. In this work, we consider
an arbitrary value of the parameter $f=<F>$ when the breaking type of the global
supersymmetry belongs to the so--called $F$--type with the $F$--term of some chiral
superfield acquiring such an expectation value.

The Lagrangian (\ref{actmg}) is characterized by a nonlinear nonholonomic
realization of global supersymetry with an infinitesimal complex\ parameter $%
\alpha $ (and its complex \ conjugate $\overline{\alpha }),$ when $\delta
\lambda =f\alpha +i\overline{\alpha }f^{-1}\lambda \gamma ^{\mu }\mathbf{e}%
_{\mu }\lambda .$ It is possible to generate a mass for the gravitino $\psi
_{\alpha }$ through the absorbtion of the Goldstino
via a super--Higgs effect. We can introduce local supersymmetry by allowing the
parameter $\alpha \rightarrow \alpha (u)$ to depend on the spacetime coordinates
$u=\{u^{\alpha }\}.$ The equation (\ref{actmg}) transforms into
\begin{equation}
\ ^{\lambda }\mathcal{L}=-f^{2}|\mathbf{e}|-\frac{i}{2}\overline{\lambda }%
\gamma ^{\mu }\mathbf{e}_{\mu }\lambda -\frac{i}{2}\overline{\lambda }\gamma
^{\mu }\psi _{\mu }+\ldots ,  \label{actmga}
\end{equation}%
where $|\mathbf{e}|=\det e_{\ \alpha }^{\underline{\alpha }}$ denotes the
determinant of the vierbein $e_{\ \alpha }^{\underline{\alpha }}$ field
which in the N--adapted form defines a decomposition of a d--metric (\ref{dm1}),
when $\mathbf{g}_{\alpha \beta }=\mathbf{e}_{\ \alpha }^{\underline{\alpha }}%
\mathbf{e}_{\ \beta }^{\underline{\beta }}\eta _{\underline{\alpha }%
\underline{\beta }}$ with $\eta _{\underline{\alpha }\underline{\beta }}$
being the Minkowski metric. This action is invariant under following
N--adapted supergravity transformations%
\begin{equation*}
\delta \lambda =f\alpha (u)+\ldots ,\ \delta e_{\ \alpha }^{\underline{%
\alpha }}=-i\kappa \overline{\alpha }(u)\gamma ^{\underline{\alpha }}\psi
_{\alpha },\ \delta \psi _{\alpha }=-2\kappa ^{-1}\mathbf{e}_{\mu }\alpha
(u)+\ldots \ .
\end{equation*}

We can impose in the N--adapted form the gauge condition $\gamma ^{\mu }\psi
_{\mu }=0,$ when the Goldstino can be gauged away by a suitable redefinition
(even in the presence of off--diagonal interactions in MGT) by
considering a suitable redefinition of the gravitino field and the vierbein
fields. Nevertheless, the term $-f^{2}|\mathbf{e}|$ results in a negative
cosmological constant.

\subsection{An Einstein--Cartan like formulation of MSGT}

We have nontrivial torsion fields which can be included in SG theories in a
form similar to that for the Einstein--Cartan theory. Nevertheless, we do
not need additional field equations for such torsion fields because they are
nonholonomically induced by the off--diagonal deformations of the metrics and the
respective vielbein fields.

\subsubsection{The nonholonomic supersymmetric Lagrangian}

Let us denote by $\widehat{\omega }_{\mu }^{\ \underline{\alpha }\underline{%
\beta }}$ the canonical\ spin d--connection determined by $\widehat{\mathbf{D%
}},$ see details \cite{vspinor} (the constructions are similar to those for $%
\omega _{\mu }^{\ \underline{\alpha }\underline{\beta }}$ constructed for $%
\nabla ).$ The so--called 1.5 order formalism of the \ Einstein--Cartan
formulation of supergravity \cite{supersymy} is extended to nonholonomic
(super) manifolds in the form%
\begin{equation}
\ ^{sg}\mathcal{L}=-\frac{|\mathbf{e}|}{2\kappa ^{2}}\vec{\mathbf{R}}-\frac{1%
}{2}\epsilon ^{\mu \nu \alpha \beta }\overline{\psi }_{\mu }\gamma
_{5}\gamma _{\nu }\widehat{\mathcal{D}}_{\alpha }\psi _{\beta }+\frac{|%
\mathbf{e}|}{3}(\mathbf{A}_{\alpha }\mathbf{A}^{\alpha }-S^{2}-P^{2}),
\label{sgact}
\end{equation}%
for $\vec{\omega}_{\mu }^{\ \underline{\alpha }\underline{\beta }}:=\widehat{%
\omega }_{\mu }^{\ \underline{\alpha }\underline{\beta }}(\mathbf{e}_{\beta
})+\frac{\kappa ^{2}}{4}(\overline{\psi }_{\mu }\gamma ^{\underline{\alpha }%
}\psi ^{\underline{\beta }}-\overline{\psi }_{\mu }\gamma ^{\underline{\beta
}}\psi ^{\underline{\alpha }}+\overline{\psi }^{\underline{\alpha }}\gamma
_{\mu }\psi ^{\underline{\beta }})$ and the covariant spin d--operator $\widehat{%
\mathcal{D}}_{\alpha }:=\mathbf{e}_{\alpha }+$ $\frac{1}{2}\widehat{\omega }%
_{\mu }^{\ \underline{\alpha }\underline{\beta }}(\mathbf{e}_{\beta })\sigma
_{\underline{\alpha }\underline{\beta }}$ defining the canonical d--spinor
curvature%
\begin{equation*}
\vec{\mathbf{R}}_{\ \ \ \mu \nu }^{\underline{\alpha }\underline{\beta }}:=%
\mathbf{e}_{\mu }\vec{\omega}_{\nu }^{\ \underline{\alpha }\underline{\beta }%
}-\mathbf{e}_{\nu }\vec{\omega}_{\mu }^{\ \underline{\alpha }\underline{%
\beta }}+\vec{\omega}_{\mu }^{\ \underline{\alpha }\underline{\tau }}\vec{%
\omega}_{\nu \underline{\tau }}^{\quad \underline{\beta }}-\vec{\omega}_{\nu
}^{\ \underline{\alpha }\underline{\tau }}\vec{\omega}_{\mu \underline{\tau }%
}^{\quad \underline{\beta }}+W_{\mu \nu }^{\gamma }\vec{\omega}_{\gamma }^{\
\underline{\alpha }\underline{\beta }},
\end{equation*}%
where $W_{\mu \nu }^{\gamma }$ are the anholonomy coefficients in (\ref%
{nonholr}). This allows us to compute the canonical scalar curvature%
\begin{equation}
\vec{\mathbf{R}}=\mathbf{e}_{\underline{\alpha }}^{\ \mu }\mathbf{e}_{%
\underline{\beta }}^{\ \nu }\vec{\mathbf{R}}_{\ \ \ \mu \nu }^{\underline{%
\alpha }\underline{\beta }}=\widehat{\mathbf{R}}(\mathbf{e}_{\beta })+\frac{%
11}{4}\kappa ^{4}(\overline{\psi }_{\mu }\Gamma ^{\mu \nu }\psi _{\nu
})^{2}+\ldots ,  \label{auxsc2}
\end{equation}%
where $\Gamma ^{\mu \nu }=\gamma ^{\lbrack \mu }\gamma ^{\nu ]}/4$ and \ the
fields $(\mathbf{A}_{\alpha },S,P)$ comprise a minimal set of auxiliary
fields which can be used for the closure of the  (super) algebra. The dots $\ldots $
indicate (as in the original Volkov--Akulov theory) the interaction terms
between the gravitino and (modified gravity fields and possible massive)
graviton fields and the four--gravitino interactions involving $\gamma _{5}.$
The SG action (\ref{sgact}) posses a local invariance under N--adapted
transforms and a large class of MGTs,%
\begin{eqnarray*}
\delta \mathbf{e}_{\ \alpha }^{\underline{\alpha }} &=&\frac{\kappa }{2}%
\overline{\epsilon }\gamma ^{\underline{\alpha }}\psi _{\alpha },~\delta
\psi _{\nu }=(\kappa ^{-1}\widehat{\mathcal{D}}_{\nu }+\frac{i}{2}\mathbf{A}%
_{\nu }\gamma ^{5}-\frac{1}{2}\psi _{\nu }\eta )\epsilon , \\
\delta S &=&\frac{1}{4}\overline{\epsilon }\gamma _{\alpha }\widehat{%
\mathcal{R}}^{\alpha },\delta P=-\frac{1}{4}\overline{\epsilon }\gamma
_{5}\gamma _{\alpha }\widehat{\mathcal{R}}^{\alpha },\delta \mathbf{A}_{\nu
}=\frac{3i}{4}\overline{\epsilon }\gamma _{5}(\widehat{\mathcal{R}}_{\nu
}-\gamma _{\nu }\gamma _{\alpha }\widehat{\mathcal{R}}^{\alpha }),
\end{eqnarray*}%
where $\epsilon $ is a complex parameter, $\eta =$ $\frac{1}{3}(-S+i\gamma
_{5}P+i\gamma ^{\alpha }\mathbf{A}_{\alpha }\gamma _{5})$ and the
'supercovariantised' and nonholonomically deformed gravitino field equation
is%
\begin{equation*}
\widehat{\mathcal{R}}^{\alpha }=\epsilon ^{\alpha \beta \mu \nu }\gamma
_{5}\gamma _{\beta }(\widehat{\mathcal{D}}_{\mu }\psi _{\nu }-\frac{i}{2}%
\mathbf{A}_{\nu }\gamma _{5}\psi _{\mu }+\frac{1}{2}\gamma _{\alpha }\eta
\psi _{\mu }).
\end{equation*}%
Explicit deformations with respect to the LC--connection can be computed for
distortions of the type in (\ref{distr}) when $\widehat{\mathbf{D}}=\nabla +%
\widehat{\mathbf{Z}}$.

\subsubsection{Off--diagonal mass generation for the gravitino}

The dynamical mass generation for the gravitino can be investigated for an
auxiliary scalar field $\sigma $ and its Euler--Lagrange equation and
linearizing the four--gravitino interactions in (\ref{auxsc2}),%
\begin{equation*}
\frac{|\mathbf{e}|}{2\kappa ^{2}}\vec{\mathbf{R}}\sim |\mathbf{e}|[\frac{1}{%
2\kappa ^{2}}\widehat{\mathbf{R}}(\mathbf{e}_{\beta })-\sigma ^{2}-\sqrt{%
\frac{11}{2}}\frac{\kappa \sigma }{2}(\overline{\psi }_{\mu }\Gamma ^{\mu
\nu }\psi _{\nu })^{2}+\ldots ].
\end{equation*}%
We obtain an effective Lagrangian with a nonholonomically Goldstino-induced
negative constant (see (\ref{actmga}))
\begin{eqnarray}
\ ^{eff}\mathcal{L} &=&-\frac{|\mathbf{e}|}{2\kappa ^{2}}[\widehat{\mathbf{R}%
}(\mathbf{e}_{\beta })+2\kappa ^{2}(f^{2}-\sigma ^{2})]
\label{actmgneutrino} \\
&&-\frac{1}{2}\epsilon ^{\mu \nu \alpha \beta }\overline{\psi }_{\mu }\gamma
_{5}\gamma _{\nu }\widehat{\mathcal{D}}_{\alpha }\psi _{\beta }+\sqrt{\frac{%
11}{2}}\frac{\kappa \sigma |\mathbf{e}|}{2}(\overline{\psi }_{\mu }\psi
^{\mu })^{2}+\ldots  \notag
\end{eqnarray}%
with distortions in $\widehat{\mathbf{R}}$ and $\widehat{\mathcal{D}}$
which for diagonal configurations in (effective) GR transform into similar
values determined by $\nabla .$ Such models with dynamical supersymmetry
breaking in LC--variables were studied in Refs. \cite{fradkin,mavr1}.
Considering that $\sigma $ may acquire a non--zero vacuum expectation value
(vev) $\sigma _{c}:=<~^{0}\sigma >+<~^{1}\sigma >\neq 0$ (this can be a
process of quantization, or via re--definition of the generating functions in
MGTs), we can obtain a dynamical breaking of local supersymmetry with a distorted effective mass for the  gravitino,%
\begin{equation}
~^{\sigma }\mu =\kappa \sqrt{\frac{11}{2}}\sigma _{c}=\kappa \sqrt{\frac{11}{%
2}}(~_{0}\sigma _{c}+~_{1}\sigma _{c})=~_{0}^{\sigma }\mu +~_{1}^{\sigma
}\mu ,  \label{sigma}
\end{equation}%
where $~_{0}^{\sigma }\mu $ is determined  from the computations with $\nabla $ and
$~_{1}^{\sigma }\mu $ $\ $\ is a \ distortion mass term determined by $%
\widehat{\mathbf{D}}.$ The terms $~_{1}\sigma _{c}$ and $~_{1}^{\sigma }\mu $
vanish for holonomic configurations and diagonal Einstein metrics.

Let us analyze the issue of the cosmological constant. We can express the above
effective Lagrangian in the form
\begin{equation}
\ ^{eff}\mathcal{L}=-\frac{|\mathbf{e}|}{2\kappa ^{2}}[\widehat{\mathbf{R}}(%
\mathbf{e}_{\beta })-2\Lambda _{0}+\check{\Lambda}]+\ldots  \label{sgact1}
\end{equation}%
where $\Lambda _{0}+\check{\Lambda}:=\kappa ^{2}(\sigma _{c}^{2}-f^{2})$ can
be identified with a tree--level cosmological constant. For off--diagonal
configurations in GR and/or MGTs, we have additional possibilities to
consider when $\kappa ^{2}(~^{1}\sigma _{c})=\check{\Lambda},$ see (\ref%
{efes}), if we do not involve quantum corrections. We conclude that even the
de Sitter spacetime may not be included into the SG effective Lagrangian $\
^{sg}\mathcal{L}$ (\ref{sgact}), such an effective constant may have two
sources: It may be a solution of a quantum effective action (including
fluctuations of the metric and gravitino and ghost fields) and/or as a result
of nonholonomic off--diagonal interactions of the gravitational and (effective)
matter fields.

\section{One--Loop Partition Functions with MGT Deformations}

\label{s4}

Let us explain how the method of computing the one--loop effective action
elaborated in section IV of \cite{mavr1} and which can be generalized to
include off--diagonal terms with the possibility of encoding the MGTs effects. There are
three key issues: 1) Expanding about a classical background \ $\mathbf{g}%
_{\mu \nu }$ (it can be a standard Euclidean $dS_{4}$ metric), we can study
small off--diagonal deformations $\mathbf{\tilde{g}}_{\mu \nu }\rightarrow
\mathbf{g}_{\mu \nu }+\zeta _{\mu \nu }$ considering that such deformations
are computed both as small noholonomic deformations and constrained
fluctuations of \ the action up to quadratic order while working with a small $%
\zeta _{\mu \nu } $ in a one--loop formalism. 2) The off--diagonal
contributions caan also be included in the distortion tensors $\widehat{\mathbf{Z}}%
. $ 3) Finally, a very important mechanism that takes into account possible MGT
effects is to consider the effective cosmological constants determined by
re--definitions of the generating functions.

Our strategy is to decompose all the physical \ and ghost fields defined on small
off--diagonally deformed background geometries and work with the data \ $(\zeta
_{\mu \nu },\widehat{\mathbf{D}},\Lambda _{0}+\check{\Lambda}).$ In the
presence of \ fermions, we shall use a N--adapted \ vierbein formalism and
the canonical spin d--connection $\widehat{\mathcal{D}}$ which should reproduce
the same results as in  the metric formalism (see\ similar constructions in
\cite{percacci} but for LC configurations). We assume also that the
one--loop effective action encoding the MGTs contributions can obtained by
expanding about a de Sitter background with a renormalized cosmological
constant $\Lambda >0$ and to consider \ that in the limit $\Lambda \rightarrow 0$
the gravitino mass reaches its physical value. The
relation between $\Lambda _{0},\check{\Lambda}$ and $\Lambda $ will be
studied \ below and which will \ impose certain conditions on the generating
parameter $\check{\Lambda}.$

\subsection{ Nonholonomic bosonic configurations}

The MGTs contributions will be encoded into the "boldface" operators when the
coefficients of the geometric objects are computed with respect to the N--adapted
frames.

\subsubsection{Bosonic operators}

Introducing $\zeta _{\mu \nu }=\overline{\zeta }_{\mu \nu }+\mathbf{g}_{\mu
\nu }\zeta /4,$ for $\zeta =\mathbf{g}^{\mu \nu }\zeta _{\mu \nu },$ we can
apply to eq-(\ref{sgact1}) the variational formalism elaborated in \cite%
{fradkin,mavr1}. Up to quadratic order in $\zeta _{\mu \nu },$ we obtain%
\footnote{%
we omit details on the second variations in the N--adapted formalism etc; the
formulas for the transformation of the N--adapted bases into the coordinate-frame ones,  when the
off--diagonal decompositions are small} the effective action {\small
\begin{equation*}
\frac{1}{4\kappa ^{2}}\int d^{4}u\sqrt{|\mathbf{g}|}[\frac{1}{2}\overline{%
\zeta }_{\mu \nu }(-\widehat{\mathbf{D}}_{\mu }\widehat{\mathbf{D}}^{\mu
}+~^{1}\lambda )\overline{\zeta }^{\mu \nu }-\frac{1}{2}\zeta (-\widehat{%
\mathbf{D}}_{\mu }\widehat{\mathbf{D}}^{\mu }+~^{2}\lambda )\zeta -(\widehat{%
\mathbf{D}}^{\mu }\overline{\zeta }_{\mu \nu }-\frac{1}{4}\widehat{\mathbf{D}%
}_{\nu })^{2}\zeta ]
\end{equation*}%
}where%
\begin{equation}
~^{1}\lambda =\frac{13}{2}\Lambda -\frac{3}{2}(\Lambda _{0}+\check{\Lambda})%
\mbox{ and }~^{2}\lambda =\frac{5}{2}(\Lambda _{0}+\check{\Lambda})-\frac{1}{%
2}\Lambda .  \label{lambda}
\end{equation}

Our goal is to compute the spectra of the three nonholonomic bosonic operators:%
\begin{eqnarray}
~^{0}\widehat{\bigtriangleup }(X)\varphi &=&(-\widehat{\mathbf{D}}_{\mu }%
\widehat{\mathbf{D}}^{\mu }+X)\varphi ,~^{1}\widehat{\bigtriangleup }^{\mu
\nu }(X)\xi _{\nu }^{\perp }=(-\widehat{\mathbf{D}}^{\mu }{}^{\nu }+\mathbf{g%
}^{\mu \nu }X)\xi _{\nu }^{\perp },  \notag \\
~^{2}\widehat{\bigtriangleup }_{\alpha \beta }^{\mu \nu }(X)\overline{\zeta }%
_{\mu \nu }^{\perp } &=&(-\widehat{\mathbf{D}}^{\mu }{}_{\alpha \beta }^{\nu
}+\delta _{\alpha }^{\mu }\delta _{\beta }^{\nu }X)\overline{\zeta }_{\mu
\nu }^{\perp },  \label{nbosop}
\end{eqnarray}%
were the operators $\widehat{\mathbf{D}}^{\mu }{}^{\nu }$ and $\widehat{%
\mathbf{D}}^{\mu }{}_{\alpha \beta }^{\nu }$ are constructed in
a similar form  \ to those presented in \cite{fradkin} but for $\nabla ^{\mu
}\rightarrow \widehat{\mathbf{D}}^{\mu }.$ In above formulas, $X$ is a
constant and the "transverse traceless' decompositions are performed in the
form:%
\begin{eqnarray}
\mathbf{Y}_{\mu } &=&\mathbf{Y}_{\mu }^{\perp }+\widehat{\mathbf{D}}_{\mu
}\varphi ,\widehat{\mathbf{D}}^{\mu }\mathbf{Y}_{\mu }^{\perp }=0,\mathcal{D}%
V=\mathcal{D}V^{\perp }\mathcal{D}\varphi \sqrt{\det ~^{0}\widehat{%
\bigtriangleup }(0)},  \label{aux51} \\
\mathbf{g}^{\mu \nu }\overline{\zeta }_{\mu \nu } &=&0,\widehat{\mathbf{D}}%
^{\mu }\overline{\zeta }_{\mu \nu }^{\perp }=0,\overline{\zeta }_{\mu \nu }=%
\overline{\zeta }_{\mu \nu }^{\perp }+2\widehat{\mathbf{D}}_{(\mu }\xi _{\nu
)}^{\perp }+\widehat{\mathbf{D}}_{\mu \nu }\chi -\frac{1}{4}\mathbf{g}_{\mu
\nu }\widehat{\mathbf{D}}_{\mu }\widehat{\mathbf{D}}^{\mu }\chi ,  \notag \\
\mathcal{D}\overline{\zeta }_{\mu \nu } &=&\mathcal{D}\overline{\zeta }%
^{\perp }\mathcal{D}\xi _{\nu }^{\perp }\mathcal{D}\chi \lbrack \det ~^{1}%
\widehat{\bigtriangleup }^{\mu \nu }(-\Lambda )]^{1/2}\otimes ~^{0}\widehat{%
\bigtriangleup }(-\frac{4}{3}\Lambda )\otimes ~^{0}\widehat{\bigtriangleup }%
(0).  \notag
\end{eqnarray}

Introducing the distortions $\widehat{\mathbf{D}}=\nabla +\widehat{\mathbf{Z}}$
in the above formulas, we can compute the distortions of operators in a similar fashion as
the ones for the  LC--configurations. For instance, $~^{0}\widehat{%
\bigtriangleup }=~_{\nabla }^{0}\bigtriangleup +$ $~_{Z}^{0}\widehat{%
\bigtriangleup }$ etc.

\subsubsection{N--adapted gauge fixing and physical gauge}

If an N--connection splitting is prescribed, both operators $\widehat{\mathbf{%
D}}$ and $\nabla $ can be used to preserve the symmetries of local Lorentz
and infinitesimal coordinate transformations. The first type of symmetries
can be fixed \ by setting the antisymmetric part of N--elongated vielbein to
be zero (such a procedure is used in \cite{supersymy}). Adding the standard
two parameters, $\alpha $ and $\beta , a $ d--covariant gauge fixing term, we can
fix the coordinate gauge transformations for the MGTs off--diagonal deformations that are
encoded into the bosonic part of the action $S_{B}$ (we write $gf$ \ as a left
label for gauge fixing),%
\begin{equation}
~^{gf}S_{B}=-\frac{1}{4\kappa ^{2}\alpha }\int d^{4}u\sqrt{|\mathbf{g}|}(%
\widehat{\mathbf{D}}^{\mu }\zeta _{\mu \nu }-\frac{1+\beta }{4}\widehat{%
\mathbf{D}}_{\nu }\zeta ).  \label{gaugf}
\end{equation}%
It is necessary to introduce a nonholonomic ghost (gh) action for some
anti--commuting complex d--vector field $\mathbf{C}^{\mu }$ when
\begin{equation*}
^{gh}S_{B}=-\frac{1}{4\kappa ^{2}\alpha }\int d^{4}u\sqrt{|\mathbf{g}|}%
\overline{\mathbf{C}}^{\mu }[-(\widehat{\mathbf{D}}_{\mu }\widehat{\mathbf{D}%
}^{\mu }+\Lambda )\delta _{\mu \nu }+\frac{\beta -1}{4}(\widehat{\mathbf{D}}%
_{\mu }\widehat{\mathbf{D}}_{\nu }+\widehat{\mathbf{D}}_{\nu }\widehat{%
\mathbf{D}}_{\mu })]\mathbf{C}^{\nu }.
\end{equation*}%
Integrating this formula and absorbing pre--factors into the normalization
of the functional measure, we obtain the ghost partition function,
\begin{equation*}
^{gh}\mathbb{Z}_{B}=\det ~^{1}\widehat{\bigtriangleup }^{\mu \nu }(-\Lambda
)\otimes ~^{0}\widehat{\bigtriangleup }(\frac{4}{\beta -3}\Lambda ).
\end{equation*}

The issue of gauge fixing is addressed in the literature in an alternative way
via a "physical" gauge. Let us briefly show how such constructions can be
performed using $\widehat{\mathbf{D}}$ instead of $\nabla .$ Considering $%
\alpha \rightarrow 0$ in (\ref{gaugf}), we have to impose the condition
\begin{equation*}
\widehat{\mathbf{D}}_{\mu }\zeta ^{\mu \nu }-\frac{1+\beta }{4}\widehat{%
\mathbf{D}}^{\nu }\zeta =0.
\end{equation*}%
Using the projections (\ref{aux51}) for $\beta =0,$ such conditions can be
written as a Killing condition,%
\begin{equation*}
\widehat{\mathbf{D}}_{\mu }\widehat{\mathbf{D}}^{\mu }\xi _{\nu }+\widehat{%
\mathbf{D}}_{\mu }\widehat{\mathbf{D}}_{\nu }\xi ^{\mu }=0
\end{equation*}%
for $\xi _{\nu }=\xi _{\nu }^{T}+\widehat{\mathbf{D}}_{\mu }\chi .$ Hence, when one incoporates
the conditions associated with the MGTs effects, we can impose the 'physical' gauge conditions $%
\xi _{\nu }=0.$

\subsubsection{MGT partition function}

We sahll follow a general--gauge calculation with the d--operators (\ref{nbosop}).
Similarly to formula (37) in \cite{mavr1},\footnote{%
we follow a different system of notations} we express the scalar part of the
action determined by $\widehat{\mathbf{D}}$ in matrix form,
\begin{equation*}
\frac{1}{4\kappa ^{2}}\int d^{4}u\sqrt{|\mathbf{g}|}\left[ \left(
\begin{array}{cc}
\zeta & \chi%
\end{array}%
\right) \left(
\begin{array}{cc}
\mathbf{A}_{1} & \mathbf{B} \\
\mathbf{B} & \mathbf{A}_{2}%
\end{array}%
\right) \left(
\begin{array}{c}
\zeta \\
\chi%
\end{array}%
\right) \right] ,
\end{equation*}%
where boldface matrix elements are used for geometric d--objects with a
canonical distortion. Integrating with respect to the fields, and up to an irrelevant
multiplicative factor, we find the bosonic partition function encoding the
MGT,%
\begin{eqnarray}
\mathbb{Z}_{B} &=&~^{gh}\mathbb{Z}_{B}\otimes \left( \frac{\det ~^{0}%
\widehat{\bigtriangleup }(-\frac{4}{3}\Lambda )\otimes ~^{0}\widehat{%
\bigtriangleup }(0)}{\det ~^{2}\widehat{\bigtriangleup }[~^{1}\lambda
]\otimes ~^{1}\widehat{\bigtriangleup }[\alpha (\frac{2}{3}\Lambda
-~^{1}\lambda )-\Lambda ]\otimes (\mathbf{A}_{1}\mathbf{A}_{2}-\mathbf{B}%
^{2})}\right) ^{1/2}  \notag \\
&=&\det ~^{1}\widehat{\bigtriangleup }[-\Lambda ]\otimes ~^{0}\widehat{%
\bigtriangleup }(\frac{4}{\beta -3}\Lambda )\otimes  \label{bpfdef} \\
&&\left[ ~^{2}\widehat{\bigtriangleup }[~^{1}\lambda ]\otimes ~^{1}\widehat{%
\bigtriangleup }\left( \alpha (\frac{2}{3}\Lambda -~^{1}\lambda )-\Lambda
\right) \otimes ~^{0}\widehat{\bigtriangleup }\left( \frac{A_{3}\pm \sqrt{%
A_{4}}}{6(\beta -3)^{2}}\right) \right] ^{-1/2}.  \notag
\end{eqnarray}%
All values in the expression for the $\mathbb{Z}_{B}$ are expressed in terms of d--operators and
algebraic combinations of $\alpha ,\beta $ and $~^{1}\lambda $ and $%
~^{2}\lambda ,$ see (\ref{lambda}),%
\begin{eqnarray*}
\mathbf{B} &\mathbf{=-}&\frac{3}{16}(1+\frac{\beta }{\alpha })~^{0}\widehat{%
\bigtriangleup }(0)~^{0}\widehat{\bigtriangleup }(-\frac{4}{3}\Lambda ),%
\mathbf{A}_{1}=\frac{1}{16}[(3+\frac{\beta ^{2}}{\alpha })\widehat{\mathbf{D}%
}_{\mu }\widehat{\mathbf{D}}^{\mu }+2~^{2}\lambda ], \\
\mathbf{A}_{2} &=&\mathbf{-}\frac{3}{16}(1+\frac{3}{\alpha })~^{0}\widehat{%
\bigtriangleup }(0)~^{0}\widehat{\bigtriangleup }(-\frac{4}{3}\Lambda )~^{0}%
\widehat{\bigtriangleup }\left( \frac{4(\alpha -3)\Lambda -6\alpha
~^{1}\lambda }{3(\alpha +3)}\right) , \\
A_{3} &=&4(6\alpha +\beta ^{2}+6\beta -9)\Lambda -6(3\alpha +\beta
^{2})~^{1}\lambda +6(\alpha +3)~^{2}\lambda , \\
A_{4} &=&4\left[ 2\left( 6\alpha +\beta (\beta +6)-9\right) \Lambda
-3(3\alpha +\beta ^{2})~^{1}\lambda +3(\alpha +3)~^{2}\lambda \right] ^{2} \\
&&+48(\beta -3)^{2}\left[ 3\alpha ~^{1}\lambda -2(\alpha -3)\Lambda \right]
~^{2}\lambda .
\end{eqnarray*}

The distorted partition function $\mathbb{Z}_{B}$ (\ref{bpfdef}) transforms
into a similar one and associated to Einstein gravity in the
Landau--DeWitt gauge if $\widehat{\mathbf{D}}\rightarrow \nabla .$ This
emphasizes the universality of the methods elaborated in \cite{fradkin,buch,mavr1}%
. We can work with different metric compatible linear connections defined by
the same metric structure via nonholonomic frame transformations and induced
torsion. Hereafter, we shall omit detailed proofs of the formulas and solutions
to the equations if they can be obtained via distortion relations of the
LC--connection $\nabla $ and with respective substitutions for $\widehat{\mathbf{D}%
}$ in the N--adapted frame calculations.

\subsection{Nonholonomic fermionic configurations}

Let us consider the operator $\widehat{\mbox{\DH}},$ when%
\begin{equation*}
-\widehat{\mbox{\DH}}^{2}\rightarrow -\widehat{\mathcal{D}}^{2}+\ \widehat{%
\mathbf{R}}/4...
\end{equation*}%
defines the "square" the covariant spin operator $\widehat{\mathcal{D}}$ with
$\widehat{\mathbf{R}}$ (\ref{sdcurv}) being the d--curvature scalar. We
can perform an Euclideanisation procedure via the transforms $\gamma ^{4}\rightarrow i\gamma
_{E}^{4},\gamma ^{\underline{j}}\rightarrow \gamma _{E}^{\underline{j}},%
\mathbf{e}^{4}\rightarrow \mathbf{e}_{E}^{4},\mathbf{e}^{\underline{j}%
}\rightarrow i\mathbf{e}_{E}^{\underline{j}},$ for $\underline{j}=1,2,3$,
and obtain a Laplace operator induced by $\widehat{\mathbf{D}}$, when $%
\widehat{\mbox{\DH}}\rightarrow i$ $\widehat{\mbox{\DH}}_{E},$ i.e. $%
\widehat{\mbox{\DH}}_{E}=-\widehat{\mathcal{D}}^{2}+\ \widehat{\mathbf{R}}%
/4. $ Considering $\psi =0$ in the gauge $\psi _{\mu }\gamma ^{\mu }=0,$ we
introduce the standard decompositions%
\begin{eqnarray*}
\psi _{\mu } &=&\varphi _{\mu }+\frac{1}{4}\gamma _{\mu }\psi ,\varphi _{\mu
}=\varphi _{\mu }^{\perp }+(\widehat{\mathcal{D}}_{\mu }-\frac{1}{4}\gamma
_{\mu }\widehat{\mbox{\DH}})\varsigma , \\
\widehat{\mathcal{D}}^{\mu }\varphi _{\mu }^{\perp } &=&0,\widehat{\mathcal{D%
}}\psi _{\mu }=\widehat{\mathcal{D}}^{\mu }\varphi _{\mu }^{\perp }\widehat{%
\mathcal{D}}\psi \widehat{\mathcal{D}}\varsigma /\sqrt{\det ~^{1/2}\widehat{%
\bigtriangleup }(-\frac{4}{3}\Lambda )},
\end{eqnarray*}%
and define the fermionic d--operators for constant $X,$%
\begin{equation}
~^{1/2}\widehat{\bigtriangleup }(X)\psi =(-\widehat{\mathcal{D}}^{2}+\Lambda
+X)\psi ,\ ~^{3/2}\widehat{\bigtriangleup }(X)\varphi _{\mu }^{\perp }=(-%
\widehat{\mathcal{D}}^{2\mu \nu }+\frac{4}{3}\Lambda \mathbf{g}^{\mu \nu }+X%
\mathbf{g}^{\mu \nu })\varphi _{\mu }^{\perp }.  \label{fdoper}
\end{equation}%
The last two terms in $\ ^{eff}\mathcal{L}$ (\ref{actmgneutrino}) can be
expressed via $(S^{4})$ the use of these d--operators and a generic mass term $%
M,${\small
\begin{equation*}
\frac{1}{2}\overline{\varphi }_{\mu }^{\perp }(\widehat{\mbox{\DH}}-M)%
\overline{\varphi }^{\perp \mu }+\frac{3}{16}\left(
\begin{array}{cc}
\overline{\varsigma } & \overline{\psi }%
\end{array}%
\right) \left(
\begin{array}{cc}
(\widehat{\mbox{\DH}}+2M)~^{1/2}\widehat{\bigtriangleup }(-\frac{4}{3}%
\Lambda ) & -~^{1/2}\widehat{\bigtriangleup }(-\frac{4}{3}\Lambda ) \\
-~^{1/2}\widehat{\bigtriangleup }(-\frac{4}{3}\Lambda ) & -(\widehat{%
\mbox{\DH}}-M)%
\end{array}%
\right) \left(
\begin{array}{c}
\varsigma \\
\psi%
\end{array}%
\right) .
\end{equation*}%
}

Applying such matrix representations and d--operators (\ref{fdoper}) (see
similar details in section IV\ B of \cite{mavr1}), we compute the gravitino
partition function,%
\begin{equation}
\mathbb{Z}_{F}=\left\{ \frac{\det ~^{3/2}\widehat{\bigtriangleup }(\frac{11}{%
2}\kappa ^{2}\sigma _{c}^{2})\otimes ~^{1/2}\widehat{\bigtriangleup }\left[
\left( \frac{3\sqrt{2}}{\sqrt{11}}\kappa \sigma _{c}-\frac{8}{11}\sqrt{-%
\frac{4}{3}\Lambda }\right) ^{2}\right] }{\left[ \det ~^{1/2}\widehat{%
\bigtriangleup }(-\frac{4}{3}\Lambda )\right] ^{3}}\right\} ^{1/4}.
\label{rpfdef}
\end{equation}%
The coefficients are parameterized as $\sigma _{c}=~_{0}\sigma
_{c}+~_{1}\sigma _{c}$ (\ref{sigma}) containing contributions from MGT in
such a form that similar results are reproduced for $\widehat{\mathbf{D%
}}\rightarrow \nabla $ and for diagonal configurations in GR.

\subsection{Effective MGT supersymmetry breaking}

We show how possible modifications of gravity theories can contribute to
supersymmetry breaking effects in more general supersymmetric models.

\subsubsection{One--loop effective potential and stability}

The distorted partition functions $\mathbb{Z}_{B}$ (\ref{bpfdef}) and $%
\mathbb{Z}_{F}$ (\ref{rpfdef}) can be used to compute the one--loop
effective action%
\begin{equation*}
~^{1loop}\mathcal{S}=-\ln (\mathbb{Z}_{B}\otimes \mathbb{Z}_{F})=\frac{1}{2}%
\ln \det ~^{2}\widehat{\bigtriangleup }[~^{1}\lambda ]+...,
\end{equation*}%
following the functional determinant techniques involving the zeta function $\zeta
(z):=\sum\nolimits_{n}^{\infty }g_{n}\lambda _{n}^{-z}$ and derivative $%
\zeta ^{\prime }(0):=-\sum\nolimits_{n}^{\infty }g_{n}\ln \lambda _{n}$ (see
a summary in the Appendix in \cite{mavr1} and Section V in that paper).  A
higher energy cut--off is considered via the short time $\epsilon =\mu
^{2}/L^{2}\rightarrow 0$ expansion with the constants $\mu $ and $L$ of mass dimension. We
can use those formulas by substituting into the functionals the respective
functions, $R\rightarrow \widehat{\mathbf{R}},\Lambda _{0}\rightarrow
\Lambda _{0}+\check{\Lambda}$ and $\sigma _{c}\rightarrow \sigma
_{c}=~_{0}\sigma _{c}+~_{1}\sigma _{c}.$ The finite part of the above functional
is expressed as
\begin{equation}
~^{1loop}\mathcal{S}=\mathcal{S}_{c}+(B_{4}-\mathcal{N})\ln (\Lambda /3\mu
^{2})-B_{4}^{\prime }  \label{1loopact}
\end{equation}%
for the effective potential
\begin{equation*}
~^{eff}\mathcal{V}=-\frac{\Lambda ^{2}}{24\pi ^{2}}~^{1loop}\mathcal{S},
\end{equation*}%
where $\mathcal{N}=14\mathcal{-}\frac{1}{2}8=10$ is the number of extra zero
modes as elucidated by \cite{fradkin} and $24\pi ^{2}/\Lambda ^{2}$ is the
spacetime volume for $S^{4}$ of radius $\sqrt{3/\Lambda }.$

Let us introduce the functions%
\begin{equation*}
\zeta _{s}(z,X)=\frac{1}{3}(2s+1)F\left[ z,2s+1,(s+1/2)^{2},b_{s}(X)\right] ,
\end{equation*}%
where $F(z,k,a,b):=\sum\nolimits_{p=k/2+1}^{\infty }\frac{p(p^{2}-a)}{%
(p^{2}-b)^{z}}$ and $b_{0}(X)=9/4-3X/\Lambda $,\newline
$b_{1/2}(X)=-3X/\Lambda ,b_{1}(X)=13/4-3X/\Lambda ,b_{3/2}(X)=-3X/\Lambda
,b_{2}(X)=17/4-3X/\Lambda $. The coefficients of the action (\ref{1loopact}) can
be computed and expressed as follows:%
\begin{eqnarray*}
\mathcal{S}_{c} &=&-\frac{1}{2\kappa ^{2}}\int d^{4}u\sqrt{|\mathbf{g}|}[%
\widehat{\mathbf{R}}-2(\Lambda _{0}+\check{\Lambda})]=-\frac{12\pi ^{2}}{%
\kappa ^{2}\Lambda ^{2}}(\Lambda _{0}+\check{\Lambda}), \\
B_{4} &=&\frac{1}{2}\zeta _{2}(0,~^{1}\lambda )-\frac{1}{4}\zeta _{3/2}(0,%
\frac{11}{2}\kappa ^{2}(~_{0}\sigma _{c}+~_{1}\sigma _{c})^{2}) \\
&&-\zeta _{1}(0,-\Lambda )+\frac{1}{2}\zeta _{1}\left[ 0,\alpha (\frac{2}{3}%
\Lambda -~^{1}\lambda )-\Lambda \right] \\
&&-\frac{1}{4}\zeta _{3/2}\left[ 0,\left( 3\sqrt{\frac{2}{11}}\kappa
(~_{0}\sigma _{c}+~_{1}\sigma _{c})-\frac{8}{11}\sqrt{-\frac{4}{3}\Lambda }%
\right) ^{2}\right] \\
&& +\frac{3}{4}\zeta _{1/2}(0,-\frac{4}{3}\Lambda ) -\zeta _{0}(0,-\frac{4}{%
\beta -3}\Lambda )-\frac{1}{2}\zeta _{0}(0,\frac{A_{3}\pm \sqrt{A_{4}}}{%
6(\beta -3)^{2}}), \\
B_{4|\Lambda \rightarrow 0}^{\prime } &\rightarrow &\frac{6s+3}{4}\frac{X^{2}%
}{\Lambda ^{2}}\left[ \frac{3}{2}-\ln (\frac{3X}{2\Lambda })\right] .
\end{eqnarray*}

For $\alpha =\beta \rightarrow 0$ and $\Lambda \rightarrow 0,$ the data $%
\mathcal{S}_{c},B_{4}$ and $B_{4|\Lambda \rightarrow 0}^{\prime }$ from the $%
~^{1loop}\mathcal{S}$ allows to compute the effective distorted potential
(the coefficients are taken to be the same as in Section V B of \cite{mavr1}, but
for additional off--diagonal contributions from the MGTs in $~_{1}\sigma _{c}$),
{\small
\begin{eqnarray*}
~^{eff}\mathcal{V} &=&f^{2}-(~_{0}\sigma _{c}+~_{1}\sigma
_{c})^{2}+\{16335f^{4}-10890[f^{2}-32670f^{2}(~_{0}\sigma _{c}+~_{1}\sigma
_{c})^{2} \\
&&-(~_{0}\sigma _{c}+~_{1}\sigma _{c})^{2}]^{2}\ln [\frac{3\kappa ^{2}}{2\mu
^{2}}(f^{2}-(~_{0}\sigma _{c}+~_{1}\sigma _{c})^{2})]+(~_{0}\sigma
_{c}+~_{1}\sigma _{c})^{4} \\
&&\lbrack 61156\ln \frac{\kappa ^{2}(~_{0}\sigma _{c}+~_{1}\sigma _{c})^{2}}{%
3\mu ^{2}}-75399+58564\ln \frac{33}{2}+2592\ln \frac{54}{11}]\}.
\end{eqnarray*}%
} The term $\ln \{\kappa ^{2}[f^{2}-(~_{0}\sigma _{c}+~_{1}\sigma
_{c})^{2}]\}=\ln [-(\Lambda _{0}+\check{\Lambda})]$ may result in an
unstable effective potential if $f^{2}<(~_{0}\sigma _{c}+~_{1}\sigma
_{c})^{2}.$ This stability condition imposes a respective limit for $%
~_{1}\sigma _{c}$ and, correspondingly, for $\check{\Lambda}$, i.e. the \
effective cosmological constant taken from the re--definition of the generating \
function is related to the stability of the \ dynamical local breaking of
supersymmetry. For certain MGTS and off--diagonal/ nonholonomic
configurations, we can obtain stable models and for other ones an
instability with an imaginary potential can be obtained.

\subsubsection{Off--diagonal super--Higgs effects and fermionic and
boson\-ic contributions}

A minimum dynamically generated gravitino mass is computed following the formula
(\ref{sigma}) when
\begin{eqnarray*}
~^{\sigma }\mu &=&\kappa \sqrt{\frac{11}{2}}(~_{0}\sigma _{c}+~_{1}\sigma
_{c})=\sqrt{\frac{11}{16\pi }}\kappa ^{2}(~_{0}\sigma _{c}+~_{1}\sigma
_{c})M_{Pl} \\
&\simeq &1.63730M_{Pl}\simeq 1.99899\times 10^{19}GeV.
\end{eqnarray*}%
The MGT contributions result in a nontrivial $~_{1}\sigma _{c}$ when the
global supersymmetry breaking scale is of the order $\sqrt{f}\simeq
0.37876M_{Pl}\simeq 4.62433\times 10^{18}GeV.$ Such nonholonomic
deformations are present at zero and one--loop order and can be separated
into \ bosonic, B, and fermionic, \ F, parts,%
\begin{equation}
~^{eff}\mathcal{V}=\mathcal{V}_{B}^{(0)}+\mathcal{V}_{B}^{(1)}+\mathcal{V}%
_{F}^{(1)},\mbox{ for }\mathcal{V}_{B}^{(0)}=-\kappa ^{-2}(\Lambda _{0}+%
\check{\Lambda}),  \label{shiggsbos}
\end{equation}%
with
\begin{eqnarray*}
\mathcal{V}_{B}^{(1)} &=&\frac{45\kappa ^{4}}{512\pi ^{2}}%
[f^{2}-(~_{0}\sigma _{c}+~_{1}\sigma _{c})^{2}]^{2}\{3-2\ln (\frac{3\kappa
^{2}}{2\mu ^{2}}[f^{2}-(~_{0}\sigma _{c}+~_{1}\sigma _{c})^{2}])\}, \\
\mathcal{V}_{F}^{(1)} &=&\frac{\kappa ^{4}(~_{0}\sigma _{c}+~_{1}\sigma
_{c})^{4}}{30976\pi ^{2}}[30578\ln \frac{\kappa ^{2}(~_{0}\sigma
_{c}+~_{1}\sigma _{c})^{2}}{3\mu ^{2}} \\
&&-45867+29282\ln \frac{33}{2}+1296\ln \frac{54}{11}].
\end{eqnarray*}

Finally, we note that for the nontrivial minima $\kappa ^{4}\mathcal{V}%
_{F}^{(1)}\simeq -0.791357$ and $\kappa ^{4}\mathcal{V}_{B}^{(1)}\simeq
0.0410402$ the tree--level cosmological constant with MGT contributions
is constrained to obey the condition
\begin{equation}
\kappa ^{2}(\Lambda _{0}+\check{\Lambda})\simeq -0.750279.  \label{cond1}
\end{equation}%
For the LC--configurations, we obtain that the one--loop cosmological
constant (vacuum energy) vanishes. This is a result of the stronger spin 2
contributions, and with the opposite sign,  the gravitino torsion
terms. Nonholonomic torsion terms naturally encode such gravitino
contributions via off--diagonal nonlinear interactions. Such a formula is similar to the one for scale symmetry breaking from total derivative densities and in the cosmological constant
problem (see \cite{guend} and references therein). It can be related to the discovery of the accelerated universe expansion and the "new cosmological constant problem" when the objective is not to explain why it is zero but, instead,  it has a very small value for the vacuum energy density. Fixing one of the values $\Lambda _{0}$ and/or $\check{\Lambda}$, in the formula (\ref{cond1}), the second one can be considered as a free parameter.

\subsubsection{On super--Higgs effects for MGT solitonic configurations}

We provided explicit examples of nonlinear solitonic gravitational waves in
MGTs in subsection \ref{ssectexe}. In general, such solutions contribute
to additional terms $~_{1}^{\sigma }\mu $ and$~_{1}\sigma _{c}$ in (\ref%
{sigma}). For certain special cases like those with generating 2-d solitonic
functions of the type $\Phi =Q(x^{2},t)s(x^{2},t)$ (\ref{2dsolit}),  we obtain zero super--Higgs effects if $\Lambda _{0}=0$.

There is a strong correlation between a quantum fluctuation value $\Lambda
_{0}$ and an effective source $\check{\Lambda}$ which fix a "scale" for the
generating functions via the formulas (\ref{aux2}) and (\ref{aux2b}). For
instance, we can chose $\check{\Lambda}$ so that $\mathcal{V}_{B}^{(0)}=-\kappa
^{-2}(\Lambda _{0}+\check{\Lambda})=0$ in (\ref{shiggsbos}) for a computed
value of $\Lambda _{0}.$ Nevertheless, it is not clear how to satisfy the
condition (\ref{cond1}) in such a case.

It seems that the super--Higgs effect is a general one which holds true for
various classes of off--diagonal configurations determined by GR and/or MGT
backgrounds. Certain distinguished backgrounds are possible because
for certain configurations $\check{\Lambda}$ has a fixed value determined by
(for instance) a configuration which defines a 3-d solitonic hierarchy like
in (\ref{solitonm1}). We can change substantially the symmetry of the background
solutions, for instance, after introducing a nontrivial $v$--conformal factor
resulting in non-Killing symmetries, see (\ref{solitonm2}). Nevertheless,
this will not change  the fixed values of $\check{\Lambda}$ and the possible quantum
fluctuations (via $\check{\omega}^{2}(u^{\alpha })$) are nonholonomically
constrained such that they result in the same $\Lambda _{0}.$

In general, we consider arbitrary solitonic configurations when the
effective KP or sine-Gordon equations are not directly involved in the
conditions to be solved by the MGT field equations. Dynamical supersymmetry
breaking effects are possible because of a nontrivial effective $\check{%
\Lambda}$ determining a model of nonholonomic Einstein manifolds with induced
torsion. But there are also nonholonomic configurations like in (\ref{solitonc}) when the generating function is fixed by the conditions that MGT field
equations are solved for a fixed value of $\check{\Lambda}.$ For general solitonic
hierarchies, we can select a subset of nonlinear waves which result in a
stable effective potential with $f^{2}>(~_{0}\sigma _{c}+~_{1}\sigma
_{c})^{2}.$ It is necessary to have an additional analysis for the "conditional"
generating functions if $f^{2}<(~_{0}\sigma _{c}+~_{1}\sigma _{c})^{2}.$ $\ $%
This can be used to classify  off--diagonal solutions
and and find if the contributions from MGT are stable or unstable.

\section{Concluding Remarks}

\label{s5}

In this article, we studied the problem how modified supergravity theories,
MSGTs, have to be formulated to be self--consistent with  various classes of modified gravity
theories, MGTs, and which were constructed  recently  in connection to modern
cosmology and quantum gravity models,  the accelerated expansion of the universe,  dark energy
and dark matter physics. We have chosen for simplicity the specific case of $%
d=4$ and $\mathcal{N}=1$ supergravity and extended the constructions for
off--diagonal configurations with generalized connections. In a certain sense,
this is a further development of the paper \cite{mavr1} and early works \cite%
{fradkin}. On alternative examples of MSGTs, see \cite{ketov}.

We have constructed one--loop effective potentials for various classes of
MGTs with nonlinear action functionals, bimetric and massive terms,
nonholonomic distributions etc,  which can be effectively encoded in
off--diagonal Einstein spaces and generalizations with nonholonomically
induced torsion. In this way a dynamical mass generation for the gravitino
field can be understood in different classes of
off-diagonal solutions. We presented certain general illustrative examples when
the off--diagonal configurations are described by nonlinear solitonic waves,
in three and two dimensions, with one/two Killing and non-Killing
symmetries.  Due to special kinds of interactions induced by the presence
of an effective nonholonomic torsion, the super Higgs effects are naturally
computed with respect to nonholonomic frames.  In such cases, the
gravitational field equations decouple in general forms, and can be integrated
also in very general forms and the various quantum fluctuations and off-diagonal
nonholonomic deformations can be computed in explicit form. It is possible
to define effective nonholonomic operators that mimic (with the use of generalized
connections and spin operators) all the calculations elaborated in \cite%
{mavr1,fradkin}.

We can take into account quantum fluctuations of the generic off--diagonal
metrics which were constructed as exact solutions in MGTs. For certain well-defined
nonholonomic constraints, the corresponding generating functions can be
re--defined  and "controlled" by an effective cosmological constant. It is
important to impose such constraints and the condition that
renormalized constant vanishes only at the end of computations. We can fix a given
scale and break the global symmetry which may be stable, or not, depending
on the type of MGTs and the chosen class of off--diagonal solutions for the
effective Einstein manifolds.

It is essential to choose such value of the effective cosmological constant,
when the imaginary parts of the effective action (having proved the
existence of the respective vacua) vanish, and attain a stable mechanism for the
acquisition of a dynamical mass for the gravitino. To fix the mass to be of the order of the
Planck mass in such supergravity models is possible,  although additional
considerations are necessary in conformal supergravities etc. A rigorous and
more complete analysis of such directions to follow and their possible applications
in modern cosmology will be the subjects of  future works.

\vskip5pt

\textbf{Acknowledgments:\ } The work is partially supported by the Program
IDEI, PN-II-ID-PCE-2011-3-0256 and by an associated visiting research
position at CERN.

\end{document}